\makeatletter
\let\DMG@origCheckCommand\CheckCommand
\def\CheckCommand{\@ifstar{\DMG@checkcmd*}{\DMG@checkcmd{}}}
\def\DMG@checkcmd#1#2{\ifx#2\showhyphens\expandafter\providecommand\else\expandafter\DMG@origCheckCommand\fi#1#2}
\makeatother
\documentclass[sigplan,10pt,nonacm]{acmart}
\usepackage{amsmath} \usepackage{graphicx}
\usepackage{textcomp} \usepackage{xcolor}
\usepackage[ruled,linesnumbered]{algorithm2e}
\newcommand{\Comment}[1]{\tcp*{#1}}
\usepackage{tabularx} 
\usepackage{xspace}
\usepackage{graphicx}
\usepackage[caption=false, font=footnotesize]{subfig}
\usepackage{tikz} \usetikzlibrary{positioning,arrows.meta}
\usepackage{pgfplots} \pgfplotsset{compat=1.18}
\usepackage{enumitem}
\usepackage{multicol}

\usepackage{microtype}          
\let\DMGMainEncodingDefault\encodingdefault
\usepackage[T1]{fontenc}
\let\encodingdefault\DMGMainEncodingDefault
\let\DMGMainRMDefault\rmdefault
\let\DMGMainSFDefault\sfdefault
\let\DMGMainTTDefault\ttdefault
\usepackage[nomath]{lmodern}
\let\rmdefault\DMGMainRMDefault
\let\sfdefault\DMGMainSFDefault
\let\ttdefault\DMGMainTTDefault
\DeclareMathVersion{rebuttalLM}
\SetSymbolFont{operators}{rebuttalLM}{OT1}{lmr}{m}{n}
\SetSymbolFont{letters}{rebuttalLM}{OML}{lmm}{m}{it}
\SetSymbolFont{symbols}{rebuttalLM}{OMS}{lmsy}{m}{n}
\SetSymbolFont{largesymbols}{rebuttalLM}{OMX}{lmex}{m}{n}
\SetSymbolFont{AMSb}{rebuttalLM}{U}{msb}{m}{n}
\SetMathAlphabet{\mathbf}{rebuttalLM}{OT1}{lmr}{bx}{n}
\SetMathAlphabet{\mathsf}{rebuttalLM}{OT1}{lmss}{m}{n}
\SetMathAlphabet{\mathit}{rebuttalLM}{OT1}{lmr}{m}{it}
\SetMathAlphabet{\mathtt}{rebuttalLM}{OT1}{lmtt}{m}{n}

\newcommand{\MGMIS}{MG-MIS\xspace}
\newcommand{\DGMIS}{DMG-MIS\xspace}
\newcommand{\REM}[1]{}
\newcommand{\RN}[1]{}

\begin{document}

\title{DMG-MIS: A Multi-GPU Algorithm for Maximal Independent Set Computation on Dynamic Graphs}
\author{Loganath Murugan}
\affiliation{%
  \institution{Department of CSE, IIT Palakkad}
  \city{Palakkad}
  \state{Kerala}
  \country{India}
}
\email{loganath011004@gmail.com}
\author{Asaduddin Ahmed}
\affiliation{%
  \institution{Department of CSE, IIT Palakkad}
  \city{Palakkad}
  \state{Kerala}
  \country{India}
}
\email{ahmedasad72425@gmail.com}
\author{Unnikrishnan Cheramangalath}
\affiliation{%
  \institution{Department of CSE, IIT Palakkad}
  \city{Palakkad}
  \state{Kerala}
  \country{India}
}
\email{unnikrishnan@iitpkd.ac.in}
\author{Rupesh Nasre}
\affiliation{%
  \institution{Department of CSE, IIT Madras}
  \city{Chennai}
  \state{Tamil Nadu}
  \country{India}
}
\email{rupesh@cse.iitm.ac.in}

\begin{abstract} 
Computing a Maximal Independent Set (MIS) is a fundamental
	problem in graph theory, critical for applications in
	scheduling, routing, and VLSI design. 
    For massive graphs
	that exceed single-GPU memory capacities, existing multi-GPU MIS frameworks
	such as MG-MIS and cuMIS \cite{17,18} are strictly designed for static graphs, making
	recomputation expensive for dynamic graphs.
    However, real-world graphs are highly dynamic,
	undergoing continuous edge insertions and deletions. 
	To address this, we present DMG-MIS, a dynamic multi-GPU algorithm designed
	to efficiently maintain a MIS under batched edge updates without requiring
	global recomputation. Our approach employs a \textit{change propagation mechanism}
	combined with \textit{localized re-executions} of Luby's algorithm. By isolating the
	cascading effects of edge mutations, DMG-MIS adjusts vertex states in
	parallel across multiple GPUs while minimizing expensive inter-GPU
	synchronization.  In addition to DMG-MIS, we introduce significant optimizations to 
	the MG-MIS baseline for static graphs. On static graphs, our optimized multi-GPU 
	baseline achieves geometric mean speedups of $\sim$19.10$\times$ on 2 GPUs and $\sim$16.91$\times$ 
	on 4 GPUs over the original MG-MIS.
	For dynamic graphs, our batched update
	scheme remains faster than recomputing from scratch for update sizes up to
	10\% of the total graph edges, while maintaining a MIS quality that
	closely matches the static recomputation. \end{abstract}

\keywords{Maximal independent sets, multi-GPU computing, Communication optimization, Dynamic graphs}

\maketitle


\section{Introduction}

Given an undirected graph $G = (V, E)$, a \textit{Maximal Independent Set}
(MIS) is a subset $S \subseteq V$ such that (i)~no two vertices in $S$ are
adjacent (independence) and (ii)~no vertex outside $S$ can be added to $S$
without violating independence (maximality). 
Unlike the NP-hard
\textit{maximum} independent set problem, which seeks the largest possible
independent set, a MIS can be found in $O(\log n)$ parallel rounds via Luby's
randomized algorithm~\cite{5}. MIS computation has applications in parallel job
scheduling~\cite{1}, frequency assignment in wireless networks~\cite{2}, graph
coloring~\cite{3}, and VLSI design~\cite{4}.

Real-world graphs -- social networks, communication topologies, road networks,
biological interaction maps -- are rarely static. They evolve continuously through
edge insertions and deletions. In a batch-dynamic setting, an algorithm
receives a sequence of update batches and must maintain a valid MIS after every
batch. Recomputing the MIS from scratch after each batch discards the already
computed MIS and incurs cost proportional to the full graph size, even when
only a small fraction of edges changes. This makes full recomputation
increasingly expensive as graphs grow in size.

\REM{Moreover, graphs beyond a size do not fit in a single GPU's memory.}
Large graphs do not fit in the memory of a single GPU.
Processing such graphs requires distributing them across multiple GPUs.
However, graph partitioning and distributed computation on graph introduce inter-device
communication at multiple synchronization points of the underlying algorithm. It becomes imperative to control this communication cost to achieve better throughput.

Although parallel batch-dynamic MIS has been theoretically proposed~\cite{12}, and practical dynamic MIS algorithms have recently been developed for multicore CPUs and single GPUs~\cite{14,36,37}, existing GPU approaches assume that the entire graph and auxiliary maintenance structures fit within a single GPU's memory. To the best of our knowledge, multi-GPU MIS computation remains limited to the static setting in the literature~\cite{17,18}. Our proposed DMG-MIS is the first work that maintains a valid MIS under batched edge updates across multiple GPUs.

An important aspect of a parallel dynamic MIS algorithm is that a vertex originally marked to be a part of a MIS may not remain so after an update. This poses a challenge that such a conversion may trigger a  propagation of changes in the graph, starting from the converted vertex. This is expensive. Additionally, in the multi-GPU setting,  propagation of changes can nullify the MIS computation done earlier on a device, leading to additional overheads. It would be ideal if a multi-GPU MIS performs local computation on each GPU and then \textit{intelligently} combines these partial MISs making minor adjustments along the boundary. Our DMG-MIS addresses this issue.

In summary, this paper makes the following contributions: \begin{itemize} 
\item
		We present DMG-MIS, a novel dynamic MIS algorithm based on Luby's algorithm which maintains
	a valid MIS under batched edge insertions and deletions. 
\item DMG-MIS is the first multi-GPU dynamic MIS algorithm and devises local computation followed by efficient synchronization, to optimize the MIS processing. 
\item Our new ideas also lead to
an optimized \textit{static} multi-GPU MIS algorithm that eliminates the inter-device
synchronization bottleneck of the static multi-GPU MIS algorithm (MG-MIS).
\item Using a suite of nine graphs with up to a billion edges, we illustrate the effectiveness of DMG-MIS over a 4-GPU setup, achieving a speedup of 16.91$\times$ over the static multi-GPU baseline with over 99\% MIS match. Our ablation study also shows effectiveness of the proposed algorithm across varying batch sizes.
\end{itemize}
The rest of the paper is organized as follows. The section~\ref{sec:related} covers the related work. The section~\ref{sec:motivation} covers the preliminaries and motivation and the section~\ref{sec:overview} covers our algorithm and its implementation. The section~\ref{sec:expt} shows the experimental results and we conclude in section~\ref{sec:conclusion}.

\section{Related Work}\label{sec:related}
\subsection{Sequential and Parallel Dynamic MIS}
Maintaining an MIS under edge insertions and deletions has been studied
extensively in the sequential setting. Assadi et al.~\cite{6} gave the first
algorithm with update time sublinear in the number of edges, achieving
$O(m^{3/4})$ amortized update time deterministically. Gupta and Khan~\cite{8}
(and independently Du and Zhang \cite{29}) improved this to $\tilde{O}(m^{2/3})$ using
deterministic schemes, and Assadi et al.~\cite{7} subsequently achieved
$\tilde{O}(\min\{\sqrt{n}, m^{1/3}\})$ expected amortized update time using
randomized techniques against an oblivious adversary. Chechik and Zhang~\cite{9} and Behnezhad et al.~\cite{11} then independently broke this
polynomial barrier, giving the first polylogarithmic-time algorithms for the
problem: both maintain a lexicographically first MIS (LFMIS) under a uniformly
random vertex ordering. Chechik and Zhang attain $O(\log^4 n)$ expected
worst-case update time, boostable via the deamortization technique of
Bernstein, Forster, and Henzinger~\cite{10} to $O(\log^6 n)$ worst-case update
time with high probability; Behnezhad et al. attain $O(\log^2\Delta\cdot\log^2
n)$ expected time, which can likewise be adjusted to $O(\log^2\Delta\cdot\log^4
n)$ worst-case update time with high probability, where $\Delta$ is the maximum vertex
degree. Both the analyses draw on a locality property established by Censor-Hillel,
et al.~\cite{13} for the distributed dynamic MIS problem: under a
uniformly random vertex order, the expected number of vertices that must change
their MIS membership after a single edge or vertex update is at most 1.\footnote{m and n denote number of edges and vertices in the graph}
\RN{[13] mention optimal in the title. Can that be brought into our text?}

Blelloch et al.~\cite{12} extended these results to the \textit{parallel
batch-dynamic} setting, processing a batch of $b$ updates with $O(b \log^3 n)$
expected work and polylogarithmic depth. While this is an important theoretical
advance, it assumes a single shared address space and does not address
multi-GPU systems where the graph is physically partitioned across devices and
inter-device communication is orders of magnitude more expensive than a local
memory access. In principle, the approach of~\cite{12} could be adapted to a multi-GPU setting; however, their algorithm relies on fine-grained concurrent hash tables and work-stealing schedulers that require frequent random accesses across the full vertex set. On a partitioned multi-GPU system, each such access to a remote vertex incurs a PCIe transfer, making the practical overhead prohibitive compared to our approach, which confines re-evaluation to local partitions and synchronizes only the boundary state changes.
Note that all of these works have focused on the theoretical results.

\subsection{MIS Computation}

\REM{GPU implementations of \textit{static} MIS have matured considerably. ECL-MIS~\cite{15}
computes MIS using a priority-based greedy approach where vertex priorities
derive from degree and a hash-based tiebreaker, and vertices join the MIS
asynchronously without global barriers. More recently, Nijhara and Banerjee~\cite{16} proposed TC-MIS, which reformulates key MIS phases as sparse
matrix-vector multiplication and executes them on tensor cores via WMMA
instructions, achieving up to $44\times$ speedup over ECL-MIS on Hopper-class
GPUs.

For \textit{dynamic} MIS on GPUs, Nijhara et al.~\cite{14}
proposed FastMIS, which handles batched edge insertions and deletions by
classifying each edge update according to the MIS membership of its endpoints
and applying local corrections. FastMIS is a single-GPU solution and it is unclear how to extend it for efficient multi-GPU execution.
We also observed that FastMIS can produce an approximate MIS due to race conditions when processing concurrent updates (Section~\ref{sec:motivation}).}

GPU implementations of \textit{static} MIS have matured considerably, with ECL-MIS~\cite{15} employing an asynchronous priority-based greedy approach, and TC-MIS~\cite{16} reformulating MIS phases as sparse matrix-vector multiplications for tensor cores to achieve massive speedups. For \textit{dynamic} MIS, Trivedi et al.~\cite{37} introduced an incremental parallel update scheme, while FastMIS~\cite{14} extended this to batched updates via endpoint classification and local corrections using a Tuple Valued Bitmap with Levels (TVBL). However, FastMIS targets single-type batches on CPUs and, as noted in Section~\ref{sec:motivation}, can produce an invalid MIS due to race conditions under concurrent overlapping updates. More recently, Nijhara et al.~\cite{36} proposed DyGMIS, a fully dynamic batch-parallel MIS algorithm for CPUs and single GPUs that avoids write conflicts by grouping updates into disjoint two-hop zones. However, DyGMIS assumes that the entire graph and auxiliary structures fit within a single GPU's memory. In contrast, DMG-MIS partitions graphs across multiple GPUs using faimGraph and maintains the MIS via local cover counts ($\text{level}$) and localized Luby re-evaluation.

\subsection{Multi-GPU Graph Algorithms}

\REM{Akathoott et al.~\cite{18} introduced MG-MIS, the first
multi-GPU MIS algorithm. MG-MIS partitions vertices contiguously across GPUs
and runs a priority-based Luby computation with bulk inter-device
communication. It demonstrated a geometric mean speedup of $17.73\times$ over
ECL-MIS with Unified Virtual Memory on four V100 GPUs for large graphs.
MG-MIS is a \textit{static} algorithm. Our analysis also reveals severe
synchronization bottleneck in MG-MIS (discussed in Section~\ref{sec:motivation}).

More recently, Nke et al.~\cite{17} proposed cuMIS, a unified framework
spanning single-GPU, multi-GPU, and distributed multi-node configurations.
cuMIS restricts computation to an active set of undecided vertices and uses
degree-aware workload distribution to reduce imbalance and thread divergence.
It reports speedups of up to $6.5\times$ over ECL-MIS and $156\times$ over
MG-MIS, and scales to trillion-edge graphs in distributed settings where
prior approaches fail to operate. cuMIS is also a \textit{static} algorithm
and does not support incremental maintenance under edge insertions and
deletions.

Beyond MIS, frameworks such as Gunrock \cite{30}, cuGraph \cite{31}, and
Legate \cite{32} have advanced GPU-centric graph processing, and libraries
like NCCL \cite{33} and NVSHMEM \cite{34} provide inter-GPU communication
primitives. However, none of these systems provide a dynamic MIS primitive.}

In the multi-GPU setting, MG-MIS~\cite{18} partitions vertices across devices and computes MIS via Luby's priority-based approach, achieving geometric-mean speedups of $13.72\times$, $15.86\times$, and $17.73\times$ on 2, 3, and 4 GPUs respectively over ECL-MIS-UVM (evaluated on 2~NVIDIA RTX~3080 GPUs and 4~NVIDIA V100 GPUs); however, its barrier-synchronized communication at every iteration causes severe synchronization overhead (Section~\ref{sec:motivation}). Building on this, cuMIS~\cite{17} introduces a unified single-GPU, multi-GPU, and distributed framework built specifically for \textit{static} graphs: it restricts computation to active undecided vertices with degree-aware workload distribution and scales to trillion-edge graphs with speedups up to $6.5\times$ over ECL-MIS and $156\times$ over MG-MIS. However, because cuMIS is designed around static CSR graphs with degree-based vertex renumbering and read-only priority checks, it cannot adapt incrementally to graph mutations; in contrast, DMG-MIS targets \textit{dynamic} multi-GPU graphs via faimGraph and maintains the MIS incrementally under batched edge insertions and deletions. While frameworks like Gunrock~\cite{30} and cuGraph~\cite{31} advance general GPU graph processing and libraries like NCCL~\cite{33} provide communication primitives, both MG-MIS and cuMIS remain strictly static and offer no support for dynamic edge updates.

\subsection{Dynamic Graph Data Structures}
The standard Compressed Sparse Row (CSR) format is poorly suited to dynamic graphs, since any edge modification can require shifting large contiguous regions of the array. This has motivated several GPU-resident alternatives. cuSTINGER~\cite{20} stores adjacency lists in over-allocated arrays for fast updates but suffers progressive fragmentation from repeated resizing; Hornet~\cite{21} uses power-of-two block-arrays indexed by B+ trees, bounding worst-case memory overhead to a factor of two while avoiding relocation cost; GPMA~\cite{22} maintains a globally sorted gapped array with $O(\log^2 n)$ amortized updates but can trigger cascading rebalances on large batches; and Meerkat~\cite{23} uses GPU-resident hash tables~\cite{35} for expected $O(1)$ per-edge updates, at the cost of tombstone-based deletions and no sorted-order traversal. Meerkat uses warp cooperative execution model. 

We adopt faimGraph~\cite{19} for DMG-MIS. It organizes each vertex's adjacency list as a linked sequence of fixed-size pages drawn from a centralized GPU memory pool, achieving $O(1)$ amortized update cost with bounded fragmentation. This per-vertex page-linked layout maps naturally onto our multi-GPU setting, where each device maintains an independent faimGraph instance and applies edge updates in place.
\section{Preliminaries and Motivation}\label{sec:motivation}

This section establishes the algorithmic foundations used by our \DGMIS and
motivates its design by analyzing the limitations of existing approaches.

\subsection{Preliminaries}

\REM{
Luby's \textit{randomized} parallel MIS algorithm~\cite{5} proceeds in rounds. A vertex of a graph object
can be in one of the three states at a given point of time: (i) \textit{IN}, indicating that the vertex is a member of the MIS, (ii) \textit{OUT}, indicating that the vertex is not a part of the MIS. This occurs because one or more neighbors of the vertex are in the MIS, and (iii) \textit{UNDECIDED}, indicating it is not yet resolved whether the vertex is part of the MIS or not.  
Initial state of all the vertices is
\textit{UNDECIDED}. In each round of the MIS computation, every undecided vertex
decides whether it should join the MIS. To ensure this happens without conflicts (that is, two neighbors do not simultaneously become part of the MIS), Luby's algorithm maintains priorities. Thus, initially, all the vertices are assigned random priorities and in each round, a vertex
having higher priority than all of its undecided neighbors joins the MIS.
This computation happens in parallel across all the graph vertices. Neighbors of each newly added vertex are then
excluded from consideration to MIS (since the added vertex \textit{covers} them). The algorithm terminates when no undecided
vertices remain, which happens in $O(\log n)$ rounds with high probability. 
Note that the algorithm is randomized, and can be tuned to compute an MIS deterministically.
}
Luby's parallel MIS algorithm~\cite{5} proceeds in rounds where each vertex is in one of three states: \textit{IN} (part of the MIS), \textit{OUT} (covered by an \textit{IN} neighbor), or \textit{UNDECIDED}. Let $n = |V|$ and $m = |E|$. Initially, all vertices are \textit{UNDECIDED} and assigned random priorities. In each round, any \textit{UNDECIDED} vertex with a priority higher than all of its \textit{UNDECIDED} neighbors transitions to \textit{IN}, and its neighbors transition to \textit{OUT}. This process repeats in parallel until no \textit{UNDECIDED} vertices remain, typically completing in $O(\log n)$ rounds.

\textbf{Remark.} In our implementation, priorities are assigned via a deterministic hash of the vertex ID rather than randomly, making the algorithm a priority-based greedy MIS computation (as in~\cite{1}). We continue to refer to it as ``Luby's algorithm'' for exposition, since the per-round logic is identical.

\REM{For storing the dynamic graph on the GPU, we adopt the faimGraph \cite{19} data
structure for reasons discussed in Section 2.}

\subsection{Limitations of the Existing Approaches}

We now analyze two key limitations of the existing approaches that directly
motivate the design of our \DGMIS.

\subsubsection{\MGMIS Synchronization Bottleneck}

\begin{figure}[htbp] \centering
\includegraphics[width=\columnwidth]{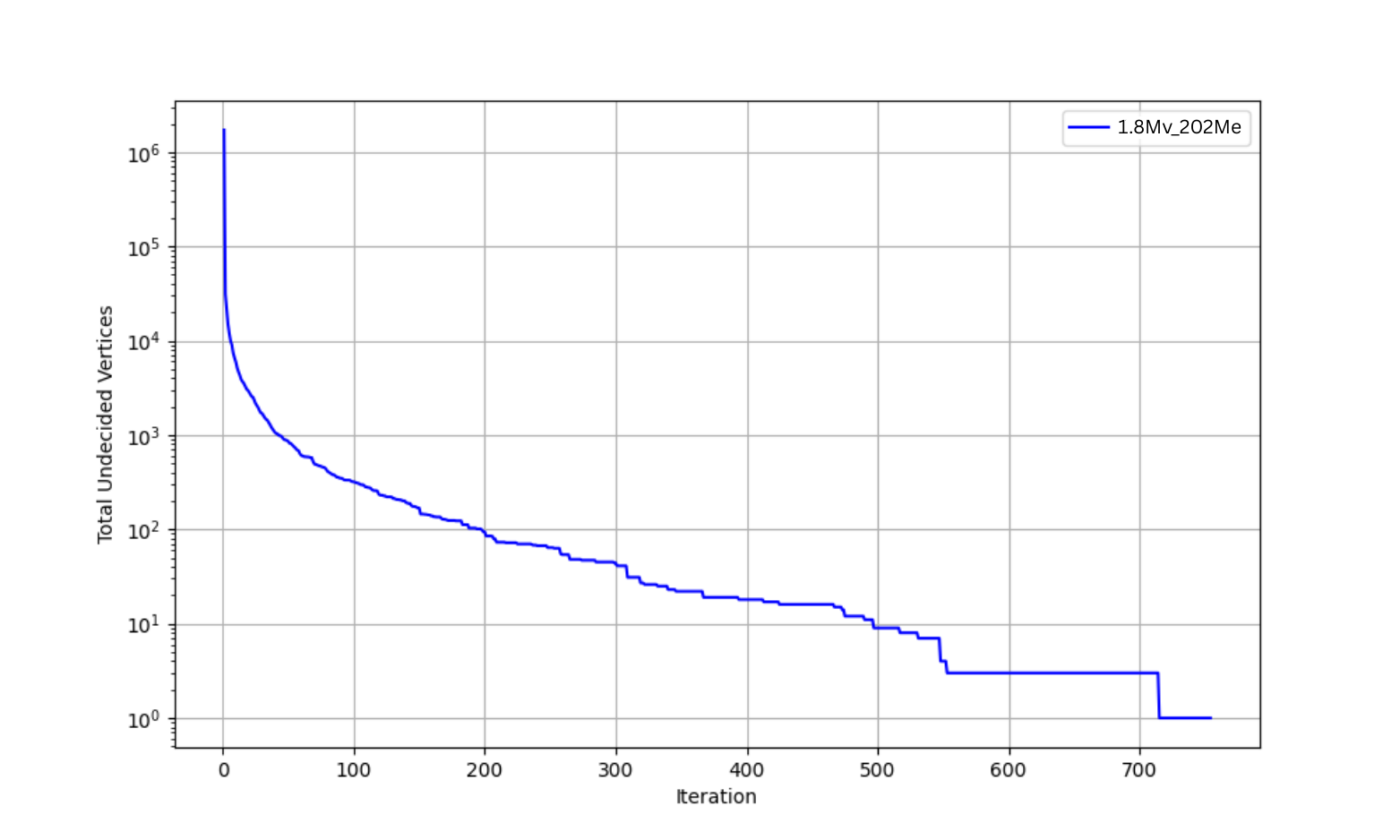}
\caption{MG-MIS convergence on a graph with 1.8M $|V|$ and 202M $|E|$.}
\Description{Plot showing MG-MIS convergence over iterations on a graph with 1.8M vertices and 202M edges, where 99 percent of vertices resolve in 10 iterations while the final 1 percent cause a 700-iteration long tail.}
\label{fig:mgmis_bottleneck} \end{figure}

\MGMIS~\cite{18} partitions vertices across GPUs and hashes vertex IDs for priorities to avoid communicating priority values. Specifically, each vertex $v$ is assigned a deterministic priority $\pi(v)$ derived from a hash of its vertex ID; ties are broken by the vertex ID itself. This allows any GPU to locally determine the priority of any vertex without communication. GPUs locally evaluate partitions in Luby rounds, synchronizing boundary states via buffers. However, \MGMIS incurs a severe synchronization bottleneck: vertices independently query remote neighbors one per iteration, without consolidating redundant queries. Thus, high-de\-gree boundary vertices require hundreds of iterations to resolve. Figure~\ref{fig:mgmis_bottleneck} shows that while 99\% of vertices resolve in 10 iterations, the final 1\% cause a 700-iter\-a\-tion long tail, severely degrading multi-GPU performance.

\begin{figure}[htbp] 
\centering
\begin{tikzpicture}[ scale=0.4, every node/.style={font=\scriptsize}, 
    node/.style={ circle, draw=black, very thick, minimum size=0.4cm, fill=white, inner sep=0.5pt }, 
    active/.style={ circle, draw=black, very thick, fill=blue!80!black, text=white, minimum size=0.4cm, inner sep=0.5pt },
    edge/.style={black, thick}, split/.style={red, dashed, very thick} ]

    \node[node, draw=none] (GPU0) at (1.5,4.5) {GPU0}; 
    \node[node, draw=none] (GPU1) at (7.5,4.5) {GPU1};
    
    \node[node] (A1) at (0,3) {$A:4$};
    \node[node] (B1) at (0,0) {$B:2$}; 
    \node[node] (C1) at (3,0) {$C:0$}; 
    
    \node[node] (D1) at (6,3) {$D:3$}; 
    \node[node] (E1) at (6,0) {$E:1$};
    \node[node] (F1) at (6,-3) {$F:5$};
    \node[node] (G1) at (9,0) {$G:6$};

    \draw[edge] (A1)--(B1); \draw[edge] (B1)--(C1);
    \draw[edge] (C1)--(D1); \draw[edge] (C1)--(E1); \draw[edge] (C1)--(F1);
    \draw[edge] (G1)--(D1); \draw[edge] (G1)--(E1); \draw[edge] (G1)--(F1);

    \draw[split] (4.5,-3.5)--(4.5,4.5); 

\end{tikzpicture}
\caption{MG-MIS working example.}
\Description{Diagram of a 7-vertex graph partitioned across GPU0 (vertices A, B, C) and GPU1 (vertices D, E, F, G) separated by a dashed boundary, with cross-GPU edges from C to D, E, and F.}
\label{fig:mgmis_example} 
\end{figure}
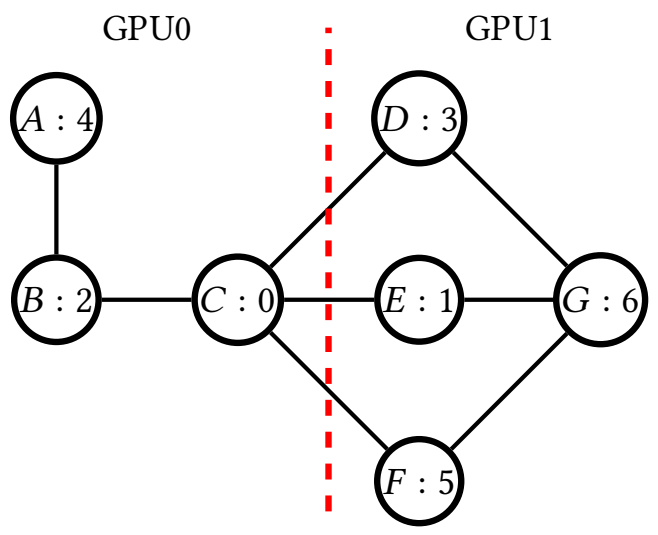 
\REM{
\subsubsection{FastMIS Race Conditions}

Nijhara, Trivedi, and Banerjee \cite{14, 17} proposed FastMIS, a dynamic MIS
algorithm utilizing a custom data structure called the Tuple Valued Bitmap with
Levels (TVBL). The TVBL structure encodes each vertex as a tuple containing a
binary bit for its MIS membership and an integer "level" representing the count
of its immediate neighbors that are currently in the MIS. When batched edge
updates arrive, FastMIS uses these decoupled state counts to perform localized
corrections rather than a full recomputation.

For edge insertions, the algorithm analyzes the state of the new edge's
endpoints. If an inserted edge connects two vertices that are both currently in
the MIS, the independence property is violated. FastMIS resolves this by
removing the lower-priority vertex from the MIS. Since this vertex is no longer
in the set, the algorithm decrements the level of all its neighbors; any
neighbor whose level reaches 0 becomes eligible to join the MIS. For edge
deletions, if a deleted edge disconnects an excluded vertex from an MIS member,
that vertex's level is decremented. If the level drops to 0, it means the
vertex no longer has any neighbors in the MIS, making it a candidate for
re-inclusion. All such vertices whose levels drop to 0 are added to the MIS
simultaneously.

While the TVBL data structure is effective for tracking localized state
changes, the FastMIS algorithm's uncoordinated approach to processing concurrent
batched updates leads to race conditions. For example, if multiple in-in conflicts occur nearby,
multiple threads will concurrently remove the lower-priority vertices and
simultaneously re-evaluate their respective neighbors without sufficient atomic
coordination. If their newly eligible neighbors are adjacent to one another,
uncoordinated re-evaluations can lead to adjacent vertices concurrently joining
the MIS, violating independence. Figure~\ref{fig:fastmis_insert} illustrates
this failure during edge insertions. Initially, vertices C, D, E, and F are in the MIS (shown in
yellow). When edges (D, F) and (C, E) are inserted concurrently, vertices C and
D are removed from the MIS, causing their respective neighbors A and B to
become eligible to join the MIS. Without coordination, both A and B join the
MIS simultaneously, violating the independence property.

\begin{figure}[!t]
\centering
\subfloat[\shortstack{Before inserting \\ edges \{(C,E),(D,F)\}})\label{fig:fastmis_insert:one}]{%
\begin{minipage}{0.48\columnwidth}
\centering
\begin{tikzpicture}[ scale=0.55, every node/.style={font=\scriptsize}, 
    node/.style={ circle, draw=black, very thick, minimum size=0.4cm, fill=white, inner sep=0.5pt }, 
    active/.style={ circle, draw=black, very thick, fill=blue!80!black, text=white, minimum size=0.4cm, inner sep=0.5pt },
    outnode/.style={ circle, draw=black, very thick, fill=gray!20, minimum size=0.4cm, inner sep=0.5pt },
    edge/.style={black, thick}, 
    newedge/.style={magenta, ultra thick, dashed} ]

    \node[outnode] (A) at (0,0) {$A:1$};
    \node[outnode] (B) at (3,0) {$B:2$};
    \node[active]  (C) at (-1,2) {$C:3$};
    \node[active]  (D) at (4,2) {$D:4$};
    \node[active]  (E) at (-1,-2) {$E:5$};
    \node[active]  (F) at (4,-2) {$F:6$};

    \draw[edge] (A)--(C);
    \draw[edge] (A)--(B);
    \draw[edge] (B)--(D);

    \draw[newedge] (C)--(E);
    \draw[newedge] (D)--(F);

\end{tikzpicture}
\end{minipage}\hfill
}
\subfloat[After insertion (A, B both join MIS —Violation)\label{fig:fastmis_insert:two}]{%
\begin{minipage}{0.48\columnwidth}
\centering
\begin{tikzpicture}[ scale=0.55, every node/.style={font=\scriptsize}, 
    node/.style={ circle, draw=black, very thick, minimum size=0.4cm, fill=white, inner sep=0.5pt }, 
    active/.style={ circle, draw=black, very thick, fill=blue!80!black, text=white, minimum size=0.4cm, inner sep=0.5pt },
    outnode/.style={ circle, draw=black, very thick, fill=gray!20, minimum size=0.4cm, inner sep=0.5pt },
    conflict/.style={ circle, draw=red, very thick, fill=red!20, minimum size=0.4cm, inner sep=0.5pt },
    edge/.style={black, thick}, 
    newedge/.style={magenta, ultra thick},
    conflictedge/.style={red, ultra thick} ]

    \node[conflict] (A) at (0,0) {$A:1$};
    \node[conflict] (B) at (3,0) {$B:2$};
    \node[outnode]  (C) at (-1,2) {$C:3$};
    \node[outnode]  (D) at (4,2) {$D:4$};
    \node[active]   (E) at (-1,-2) {$E:5$};
    \node[active]   (F) at (4,-2) {$F:6$};

    \draw[edge] (A)--(C);
    \draw[conflictedge] (A)--(B);
    \draw[edge] (B)--(D);
    \draw[newedge] (C)--(E);
    \draw[newedge] (D)--(F);

\end{tikzpicture}
\end{minipage}
}
\caption{Race condition in FastMIS during edge insertions.}
\Description{Two-panel diagram illustrating a race condition in FastMIS during concurrent insertions of edges (C,E) and (D,F), causing adjacent vertices A and B to both enter the MIS simultaneously and violate independence.}
\label{fig:fastmis_insert}
\end{figure}
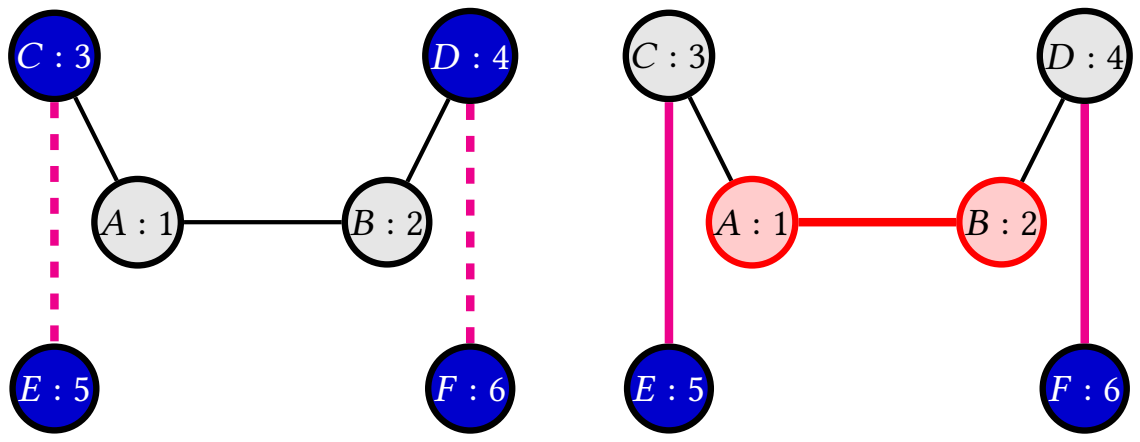

Similarly, if concurrent edge deletions cause the levels of adjacent excluded
vertices to drop to zero simultaneously, both vertices may incorrectly join the
MIS without coordinating. Figure~\ref{fig:fastmis_delete} shows this condition.
Initially, vertices C and D are in the MIS (shown in yellow). Concurrent
deletion of edges (A, D) and (B, C) leaves both A and B with no MIS neighbors.
Both transition to the MIS, again violating independence. Uncoordinated level
decrements can also lead to incorrect neighborhood counts, violating
maximality. These race conditions become increasingly prevalent as batch sizes
grow, and are particularly disruptive on scale-free graphs where hubs
participate in many concurrent updates.

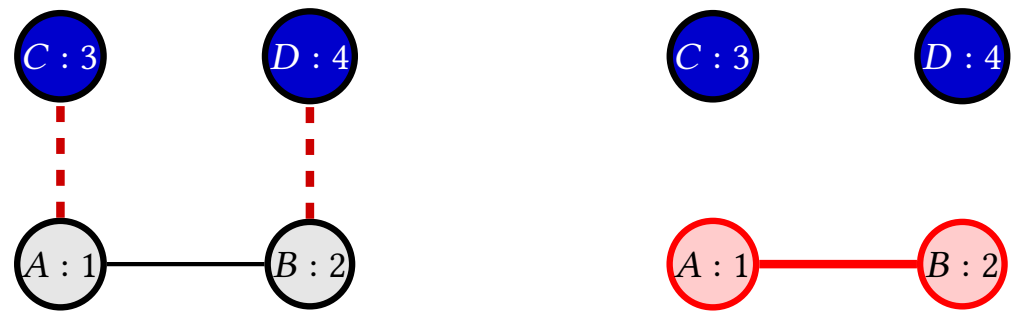
\begin{figure}[!t]
\centering
\subfloat[Before deleting edges \{(A,C),(B,D)\}]{%
\begin{minipage}{0.48\columnwidth}
\centering
\begin{tikzpicture}[ scale=0.55, every node/.style={font=\scriptsize}, 
    node/.style={ circle, draw=black, very thick, minimum size=0.4cm, fill=white, inner sep=0.5pt }, 
    active/.style={ circle, draw=black, very thick, fill=blue!80!black, text=white, minimum size=0.4cm, inner sep=0.5pt },
    outnode/.style={ circle, draw=black, very thick, fill=gray!20, minimum size=0.4cm, inner sep=0.5pt },
    edge/.style={black, thick}, 
    deledge/.style={red!80!black, ultra thick, dashed} ]

    \node[outnode] (A) at (0,0) {$A:1$};
    \node[outnode] (B) at (3,0) {$B:2$};
    \node[active]  (C) at (0,2.5) {$C:3$};
    \node[active]  (D) at (3,2.5) {$D:4$};

    \draw[edge] (A)--(B);

    \draw[deledge] (A)--(C);
    \draw[deledge] (B)--(D);

\end{tikzpicture}
\end{minipage}
}
\hfill
\subfloat[After deletion (A, B both join MIS — violation)]{%
\begin{minipage}{0.48\columnwidth}
\centering
\begin{tikzpicture}[ scale=0.55, every node/.style={font=\scriptsize}, 
    node/.style={ circle, draw=black, very thick, minimum size=0.4cm, fill=white, inner sep=0.5pt }, 
    active/.style={ circle, draw=black, very thick, fill=blue!80!black, text=white, minimum size=0.4cm, inner sep=0.5pt },
    outnode/.style={ circle, draw=black, very thick, fill=gray!20, minimum size=0.4cm, inner sep=0.5pt },
    conflict/.style={ circle, draw=red, very thick, fill=red!20, minimum size=0.4cm, inner sep=0.5pt },
    edge/.style={black, thick}, 
    conflictedge/.style={red, ultra thick} ]

    \node[conflict] (A) at (0,0) {$A:1$};
    \node[conflict] (B) at (3,0) {$B:2$};
    \node[active]   (C) at (0,2.5) {$C:3$};
    \node[active]   (D) at (3,2.5) {$D:4$};

    \draw[conflictedge] (A)--(B);

\end{tikzpicture}
\end{minipage}
}
\caption{Race condition in FastMIS during edge deletions.}
\Description{Two-panel diagram illustrating a race condition in FastMIS during concurrent deletions of edges (A,C) and (B,D), leaving adjacent vertices A and B with no MIS neighbors so both enter the MIS simultaneously and violate independence.}
\label{fig:fastmis_delete}
\end{figure}
}

\subsubsection{Working Example} 

Consider the sample graph shown in Figure ~\ref{fig:mgmis_example}\footnote{vertices denoted with vertex-id:priority ( Vertex A has priority 4)}, partitioned across two GPUs: \textit{GPU0} owns (which we refer as \textit{local vertices}) the vertices $ \{A, B, C\}$ and \textit{GPU1} owns the vertices $\{D, E, F, G\}$.

During the MIS computation, high-priority vertices $A$ and $G$ transition to \textit{IN} in the first iteration. Consequently, their neighbors $B$ (on \textit{GPU0}) and $D, E, F$ (on \textit{GPU1}) transition to \textit{OUT}. Meanwhile, $C$ on \textit{GPU0} remains \textit{UNDECIDED} and must determine the state of its remote neighbors on \textit{GPU1} to proceed.

In \MGMIS, a boundary vertex resolves remote dependencies incrementally, processing one remote query per iteration. In the first iteration, $C$ queries $D$. In the second and third iterations, it queries $E$ and $F$, respectively. Thus, \MGMIS requires three full iterations for $C$ to resolve its status as \textit{IN}, even though $D, E$, and $F$ were decided in the first round. 

\subsubsection{FastMIS Race Conditions}

Nijhara et al.~\cite{14} proposed FastMIS, which tracks vertex states and MIS neighbor counts (levels) via a Tuple Valued Bitmap with Levels (TVBL). While effective for isolated updates, uncoordinated batched updates cause race conditions.
The Figure~\ref{fig:fastmis_insert} shows the initial graph (Figure~\ref{fig:fastmis_insert:one}) and the graph after concurrent insertions of edges $(C, E)$ and $(D, F)$ (Figure~\ref{fig:fastmis_insert:two}). Resolving in-in conflicts concurrently removes lower-priority vertices (C, D) from the MIS. The adjacent neighbors (A, B) simultaneously reach level 0, they join the MIS uncoordinated, violating independence. Similarly, concurrent deletions (Figure~\ref{fig:fastmis_delete}) can simultaneously reduce adjacent excluded vertices' (A, B) levels to zero. Lacking atomic coordination, they join the MIS together. These race conditions worsen with larger batches and scale-free graphs

\begin{figure}[!t]
\centering
\subfloat[\shortstack{Before inserting \\ edges \{(C,E),(D,F)\}})\label{fig:fastmis_insert:one}]{%
\begin{minipage}{0.48\columnwidth}
\centering
\begin{tikzpicture}[ scale=0.55, every node/.style={font=\scriptsize}, 
    node/.style={ circle, draw=black, very thick, minimum size=0.4cm, fill=white, inner sep=0.5pt }, 
    active/.style={ circle, draw=black, very thick, fill=blue!80!black, text=white, minimum size=0.4cm, inner sep=0.5pt },
    outnode/.style={ circle, draw=black, very thick, fill=gray!20, minimum size=0.4cm, inner sep=0.5pt },
    edge/.style={black, thick}, 
    newedge/.style={magenta, ultra thick, dashed} ]

    \node[outnode] (A) at (0,0) {$A:1$};
    \node[outnode] (B) at (3,0) {$B:2$};
    \node[active]  (C) at (-1,2) {$C:3$};
    \node[active]  (D) at (4,2) {$D:4$};
    \node[active]  (E) at (-1,-2) {$E:5$};
    \node[active]  (F) at (4,-2) {$F:6$};

    \draw[edge] (A)--(C);
    \draw[edge] (A)--(B);
    \draw[edge] (B)--(D);

    \draw[newedge] (C)--(E);
    \draw[newedge] (D)--(F);

\end{tikzpicture}
\end{minipage}\hfill
}
\subfloat[After insertion (A, B both join MIS —Violation)\label{fig:fastmis_insert:two}]{%
\begin{minipage}{0.48\columnwidth}
\centering
\begin{tikzpicture}[ scale=0.55, every node/.style={font=\scriptsize}, 
    node/.style={ circle, draw=black, very thick, minimum size=0.4cm, fill=white, inner sep=0.5pt }, 
    active/.style={ circle, draw=black, very thick, fill=blue!80!black, text=white, minimum size=0.4cm, inner sep=0.5pt },
    outnode/.style={ circle, draw=black, very thick, fill=gray!20, minimum size=0.4cm, inner sep=0.5pt },
    conflict/.style={ circle, draw=red, very thick, fill=red!20, minimum size=0.4cm, inner sep=0.5pt },
    edge/.style={black, thick}, 
    newedge/.style={magenta, ultra thick},
    conflictedge/.style={red, ultra thick} ]

    \node[conflict] (A) at (0,0) {$A:1$};
    \node[conflict] (B) at (3,0) {$B:2$};
    \node[outnode]  (C) at (-1,2) {$C:3$};
    \node[outnode]  (D) at (4,2) {$D:4$};
    \node[active]   (E) at (-1,-2) {$E:5$};
    \node[active]   (F) at (4,-2) {$F:6$};

    \draw[edge] (A)--(C);
    \draw[conflictedge] (A)--(B);
    \draw[edge] (B)--(D);
    \draw[newedge] (C)--(E);
    \draw[newedge] (D)--(F);

\end{tikzpicture}
\end{minipage}
}
\caption{Race condition in FastMIS during edge insertions.}
\Description{Two-panel diagram illustrating a race condition in FastMIS during concurrent insertions of edges (C,E) and (D,F), causing adjacent vertices A and B to both enter the MIS simultaneously and violate independence.}
\label{fig:fastmis_insert}
\end{figure}
\begin{figure}[!t]
\centering
\subfloat[Before deleting edges \{(A,C),(B,D)\}]{%
\begin{minipage}{0.48\columnwidth}
\centering
\begin{tikzpicture}[ scale=0.55, every node/.style={font=\scriptsize}, 
    node/.style={ circle, draw=black, very thick, minimum size=0.4cm, fill=white, inner sep=0.5pt }, 
    active/.style={ circle, draw=black, very thick, fill=blue!80!black, text=white, minimum size=0.4cm, inner sep=0.5pt },
    outnode/.style={ circle, draw=black, very thick, fill=gray!20, minimum size=0.4cm, inner sep=0.5pt },
    edge/.style={black, thick}, 
    deledge/.style={red!80!black, ultra thick, dashed} ]

    \node[outnode] (A) at (0,0) {$A:1$};
    \node[outnode] (B) at (3,0) {$B:2$};
    \node[active]  (C) at (0,2.5) {$C:3$};
    \node[active]  (D) at (3,2.5) {$D:4$};

    \draw[edge] (A)--(B);

    \draw[deledge] (A)--(C);
    \draw[deledge] (B)--(D);

\end{tikzpicture}
\end{minipage}
}
\hfill
\subfloat[After deletion (A, B both join MIS — violation)]{%
\begin{minipage}{0.48\columnwidth}
\centering
\begin{tikzpicture}[ scale=0.55, every node/.style={font=\scriptsize}, 
    node/.style={ circle, draw=black, very thick, minimum size=0.4cm, fill=white, inner sep=0.5pt }, 
    active/.style={ circle, draw=black, very thick, fill=blue!80!black, text=white, minimum size=0.4cm, inner sep=0.5pt },
    outnode/.style={ circle, draw=black, very thick, fill=gray!20, minimum size=0.4cm, inner sep=0.5pt },
    conflict/.style={ circle, draw=red, very thick, fill=red!20, minimum size=0.4cm, inner sep=0.5pt },
    edge/.style={black, thick}, 
    conflictedge/.style={red, ultra thick} ]

    \node[conflict] (A) at (0,0) {$A:1$};
    \node[conflict] (B) at (3,0) {$B:2$};
    \node[active]   (C) at (0,2.5) {$C:3$};
    \node[active]   (D) at (3,2.5) {$D:4$};

    \draw[conflictedge] (A)--(B);

\end{tikzpicture}
\end{minipage}
}
\caption{Race condition in FastMIS during edge deletions.}
\Description{Two-panel diagram illustrating a race condition in FastMIS during concurrent deletions of edges (A,C) and (B,D), leaving adjacent vertices A and B with no MIS neighbors so both enter the MIS simultaneously and violate independence.}
\label{fig:fastmis_delete}
\end{figure}

\section{Core Algorithm and Implementation}\label{sec:overview}

This section extends the MIS framework to a fully dynamic distributed setting.
We define the data structures required to maintain the MIS across multiple GPUs,
the core algorithms for static MIS computation, and the procedures for handling
concurrent batched edge updates.

\subsection{Notations and Terminology} Table~\ref{tab:notations} summarizes the
key notations used throughout the algorithmic descriptions and proofs.

\begin{table}[htbp] \centering \begin{tabularx}{\columnwidth}{|l|X|} \hline
	\textbf{Notation} & \textbf{Description} \\ \hline $G = (V, E)$ &
	Undirected graph with vertex set $V$ and edge set $E$ \\ $I$ & Maximal
	Independent Set (MIS) \\ $\pi(v)$ & Priority function for vertex $v$ \\
	$N_G(v)$ & Set of neighbors of vertex $v$ in $G$ \\ $state(v)$ & State of
	vertex $v \in \{\text{IN}, \text{OUT}, \text{UNDECIDED}\}$ \\
	$\Delta E^+, \Delta E^-$ & Batch of
	concurrent edges for insertion or deletion \\ $G'$ & Updated graph after
	applying $\Delta E^+$ or $\Delta E^-$ \\ \hline \end{tabularx} \caption{Key
notations used in DMG-MIS.} \label{tab:notations} \end{table}
\subsection{Data Structures and Partitioning}
The vertex set $V$ is divided into $N$ contiguous partitions, where $N$ is
the number of devices. Partition boundaries are chosen so that each
partition's vertex count plus its incident edge count approximates to
$(|V| + |E|) / N$, giving every GPU a comparable memory footprint. We
compute these boundaries with a prefix sum over $deg(v) + 1$ for each
vertex $v$, then split the resulting cumulative sequence into $N$
contiguous segments of approximately equal total weight. Each GPU
maintains an independent faimGraph\cite{19} dynamic graph data structure instance storing only its
local partition's vertices.

Although the graph is partitioned, two state arrays, $\text{mark}$ and
$\text{level}$, are replicated at full size $|V|$ on every device, This enables
any device read the state of any vertex without querying its owning device.
After each round, devices exchange only the entries that changed, rather
than performing a full synchronization. The $\text{mark}$ array is a Boolean
array: $\texttt{true}$ if the vertex is in the MIS or \textit{UNDECIDED}, $\texttt{false}$
if excluded. The $\text{level}$ array is an integer array storing the
independent cover size $|C_G(v)|$, total count of $N_G(v)$ part of MIS. 

\begin{definition}[Independent Cover]\label{def:independent_cover}
The \emph{independent cover} of a vertex $v$ in graph $G$ with MIS $I$ is
$C_G(v) = N_G(v) \cap I$. For a vertex $v$ with $state(v) = \text{OUT}$,
$\text{level}[v] = |C_G(v)|$.
\end{definition}

Table~\ref{tab:vertex_states} summarizes how each vertex state is encoded
using the $\text{mark}$ and $\text{level}$ arrays. Vertices with $state(v) = \text{IN}$ and $state(v) = \text{UNDECIDED}$
share the boolean $\text{mark}$ state of $\text{true}$, but can be differentiated by
their $\text{level}$: vertices with $state(v) = \text{IN}$ store $\text{level} = \infty$
(represented as $\text{UINT\_MAX}$), whereas vertices with $state(v) = \text{UNDECIDED}$ have
$\text{level} = 0$.

\begin{table}[htbp] \centering \small \setlength{\tabcolsep}{3pt}
	\begin{tabularx}{\columnwidth}{|l|c|c|X|} \hline \textbf{State} &
		\textbf{mark} & \textbf{level} & \textbf{Description}
		\\ \hline IN & $\text{true}$ & $\infty (UINT\_MAX)$ & Member of the MIS. \\ 
    OUT & $\text{false}$ & $\ge 1$ & Excluded from MIS. \\ UNDECIDED & $\text{true}$ & $0$ & Not yet
	resolved. \\ \hline \end{tabularx} \caption{Vertex states and their encoding
	via the replicated $\text{mark}$ and $\text{level}$ arrays.}
	\label{tab:vertex_states} \end{table}
We employ the same priority scheme used in \MGMIS: $\pi(v)$ is a deterministic
hash of the vertex ID, with ties broken by the vertex ID itself. Because the
hash depends exclusively on the vertex ID, any device can locally compute
$\pi(v)$ for any vertex.

\subsection{Multi-GPU Luby's MIS Algorithm}

\begin{algorithm}[t]
\caption{multiGpuMis($G_i$)}
\label{algo1:luby_multi}
\SetKwProg{Fn}{}{ \{}{\}}
\Fn(){{multiGpuMis(\textup{\texttt{Graph}} $G_i$, \textup{\texttt{bool *}}mark, \textup{\texttt{int *}}level)}}{
\While{true}{
    $done\_local=\texttt{true}$;\label{algo1:luby_multi:initdlocal}
    
    MISLocal\_gpu($G_i$, $done\_local$);\label{algo1:luby_multi:mislocal}

    setToOut\_gpu();\label{algo1:luby_multi:settoout}

    syncLevel(level);\label{algo1:luby_multi:synclevel}

    syncMark(mark);\label{algo1:luby_multi:syncmark}
        
    $done\_global$ = allReduce(done\_local,\text{AND});\label{algo1:luby_multi:initdglobal}\label{algo1:luby_multi:callreduce}

    \If{$done\_global == \text{true}$}{
        \textbf{break};
    }\label{algo1:luby_multi:exitcond}
   
    }
}
\end{algorithm}

In a multi-GPU setting, the graph G is distributed across the GPUs. A vertex owned by
$GPU_i$ may have edges connecting to vertices owned by $GPU_j$, where $i\neq j$. To handle these remote edges,
the iterative algorithm (Algorithm~\ref{algo1:luby_multi}) is divided into two primary phases per
iteration (Lines~\ref{algo1:luby_multi:initdlocal}--\ref{algo1:luby_multi:syncmark}, Algorithm~\ref{algo1:luby_multi}), followed by a convergence check  (Line~\ref{algo1:luby_multi:initdglobal}-\ref{algo1:luby_multi:exitcond}, Algorithm~\ref{algo1:luby_multi}).

\paragraph{Local Computation Phase} During the local computation phase
(Lines~\ref{algo1:luby_multi:initdlocal}--~\ref{algo1:luby_multi:mislocal}), each GPU processes its partition $G_i$ concurrently. The $\textit{done\_local}$ flag is initialized to $\texttt{true}$ at the start of each iteration (Line~\ref{algo1:luby_multi:initdlocal}). The ${MISLocal\_gpu}$ kernel (Line~\ref{algo1:luby_multi:mislocal}) evaluates every local vertex $v$ with $state(v) = \text{UNDECIDED}$: (i)  vertex $v$ with one or more neighbors $u \in N_G(v)$ having $state(u) = \text{IN}$ transitions to $state(v) = \text{OUT}$ (ii)  vertex $v$ with $\pi(v)> \pi(u)~and~ state(u)==UNDECIDED~or~OUT~ \forall u ~\in ~N_G(v)$ transitions to the state $\text{IN}$ (iii) if  vertex $v$ does not satisfy condition (i) and (ii) it defers its decision, remains in  the state \text{UNDECIDED} and sets the $\text{done\_local}$ flag to $\texttt{false}$, indicating that at least one vertex on this GPU remains unresolved. 
\par The $\textit{setToOut\_gpu}$ kernel (Line~\ref{algo1:luby_multi:settoout}) then processes the vertices that had a state transition from \REM{$state(v) = $}$\text{UNDECIDED}$ to \REM{$state(v) =$} $\text{IN}$.  For neighbors of these vertices, the state is made \REM{$state(v) =$} $\text{OUT}$ and $level$ is incremented by a value of one.  To maximize memory bandwidth and prevent thread divergence, $\textit{setToOut\_gpu}$ employs a warp-cooperative traversal approach~\cite{23,38}. Rather than assigning a single thread per vertex, all threads within a warp collaboratively process the adjacency list of one active vertex at a time, ensuring fully coalesced memory accesses. Because a local vertex with \REM{$state(v) = $} $\text{IN}$ may have edges to remote vertices, $\textit{setToOut\_GPU}$ \REM{(Line~\ref{algo1:luby_multi:synclevel})}accumulates the $\text{level}$ increments for those remote vertices locally\REM{ for synchronization}.

\paragraph{Synchronization Phase} To ensure global consistency across
partitions, a two-step synchronization is performed (Lines~\ref{algo1:luby_multi:synclevel}--\ref{algo1:luby_multi:syncmark}).

\begin{itemize}
\item $\textsc{syncLevel}$ (Line~\ref{algo1:luby_multi:synclevel}): Each GPU sends the accumulated
$\text{level}$ increments for remote vertices to their respective owning GPUs. The owning GPU accumulates these updates and updates value of $\text{level}$. A vertex $v$ with $state(v) = \text{UNDECIDED}$ transitions to $state(v) = \text{OUT}$ if \REM{it receives a
positive increment for $\text{level}$} $\texttt{level(v)} \geq 1$. 

\item $\textsc{syncMark}$ (Line~\ref{algo1:luby_multi:syncmark}): Every GPU broadcasts the updated states of only those vertices whose states changed during the current iteration as $(v, \text{mark}(v))$, ensuring a globally consistent view before the next iteration.
\end{itemize}
\paragraph{Termination check}(Lines~\ref{algo1:luby_multi:initdglobal}--\ref{algo1:luby_multi:exitcond}) An $\textsc{allReduce}$ operation (Line~\ref{algo1:luby_multi:callreduce})
computes the global termination condition by performing a logical AND on the $\textit{done\_local}$ variable across all GPUs. The algorithm terminates when all vertices are globally decided, i.e \textit{done\_local} remains \texttt{true} for all GPUs after the ${MISLocal\_gpu}$ kernel execution.

\par The multiGPUMis function (Algorithm~\ref{algo1:luby_multi}) is executed from the \textit{host} (CPU)  in parallel for each subgraph $G_i$, with a separate thread $i$ for each \textit{device} ($GPU_i$). The $\textit{MISLocal\_gpu}$ (Line~\ref{algo1:luby_multi:mislocal}) and $\textit{setToOut\_gpu}$ (Line~\ref{algo1:luby_multi:settoout}) calls are GPU kernels launched on each GPU. The $\text{level}$ and $\text{mark}$ arrays are then synchronized across all GPUs via $\textit{syncLevel}$ (Line~\ref{algo1:luby_multi:synclevel}) and $\textit{syncMark}$ functions (Line~\ref{algo1:luby_multi:syncmark}), respectively.
\subsection{Dynamic MIS Rules: Handling Edge Updates} When processing
concurrent batches of edge insertions ($\Delta E^+$) or deletions ($\Delta
E^-$), the existing MIS may temporarily violate independence and maximality.
Depending on the operation, affected vertices transition to the
$state(v) = \text{UNDECIDED}$. The Luby algorithm is then re-run on this subset of
vertices with $state(v) = \text{UNDECIDED}$ to calculate a valid MIS on the updated graph $G'$.
During this re-run, only the edges among the UNDECIDED vertices in $G'$ are considered; vertices outside the UNDECIDED subset that are already OUT contribute no constraint (since they are covered by retained IN vertices), so their incident edges need not be re-examined.

\textbf{Adversary assumption.} Because priorities are deterministic (derived from a hash of vertex IDs), an adaptive adversary that observes the priority order could construct worst-case update sequences. Our analysis and experiments assume an \textit{oblivious adversary}, where the update batch is chosen independently of the priority assignment.

\subsubsection{Multi-GPU Incremental Updates}
\REM{
\begin{algorithm}[t]
\caption{multiGpuEdgeInsertions}
\label{alg:insert}
\SetKwProg{Fn}{}{ \{}{\}}
\Fn(){multiGpuEdgeInsertions(\texttt{Graph} $G_i$, \texttt{bool *}mark, \texttt{int *}level, $\Delta E^+$)}{
    \ForEach{$(src, dst) \in \Delta E^+ \mid state(dst) == \text{IN}$ \textbf{in parallel}}{
        $\text{level}[src] = \text{level}[src] + 1$;

        \If{$state(src) == \text{IN}$ \textbf{and} $\pi(dst) > \pi(src)$}{
            Flag $src$ as revoked;
        }
    }

    \ForEach{revoked vertex $w$ \textbf{in parallel}}{
        \ForEach{neighbor $x \in N_{G'}(w)$}{
            $\text{level}[x] = \text{level}[x] - 1$;
        }
    }

    syncLevel(level);

    \ForEach{local vertex $u$ \textbf{in parallel}}{
        \eIf{$u \in revoked$}{
            $\text{level}[u] = \text{level}[u] + 1$;

            \eIf{$\text{level}[u] == 0$}{
                $\text{mark}[u] = \text{UNDECIDED}$;
            }{
                $\text{mark}[u] = \text{OUT}$;
            }
        }{
            \If{$\text{mark}[u] == \text{OUT}$ \textbf{and} $\text{level}[u] == 0$}{
                $\text{mark}[u] = \text{UNDECIDED}$;
            }
        }
    }

    syncMark(mark);

    multiGpuMis($G_i$, mark, level);
}
\end{algorithm}
}
\begin{algorithm}[t]
\caption{Multi-GPU Incremental Updates}
\label{alg:insert}
\SetKwProg{Fn}{}{ \{}{\}}
\Fn(){{multiGpuEdgeInsertions(\textup{\texttt{Graph}} $G_i$, \textup{\texttt{Edges}} $\Delta E^+$,\\ \textup{\texttt{int *}} level, \textup{\texttt{bool *}} mark)}}{
    \ForEach{$(src, dst) \in \Delta E^+$ \textbf{in parallel}}{ \label{algo2:l1}
    \If{$state(dst) == \text{IN}$}{ \label{algo2:l2}
        \textbf{atomic}($\text{level}[src] = \text{level}[src] + 1$); \label{algo2:l3}

        \If{$state(src) == \text{IN}$ \textbf{and} $\pi(dst) > \pi(src)$}{ \label{algo2:l4}
            revoked[$src$] = \texttt{true}; \label{algo2:l5}
        } \label{algo2:l6}
    }
    } \label{algo2:p1end}


\tcp{cudaDeviceSynchronize()}
    
    \texttt{Sync\_With\_GPU()};

    \ForEach{local vertex $u$ \textbf{in parallel}}{ \label{algo2:l7}
        \If{revoked[$u$] == \texttt{true}}{ \label{algo2:l8}
            \ForEach{neighbor $v \in N_{G'}(u)$}{ \label{algo2:l9}
             \textbf{atomic}($\text{level}[v] = \text{level}[v] - 1$); \label{algo2:l10}
            }
        }
    } \label{algo2:l13}

    syncLevelInsert(level); \label{algo2:l14}

    \ForEach{local vertex $u$ \textbf{in parallel}}{ \label{algo2:l15}
        \eIf{revoked[$u$] == \texttt{true}}{ \label{algo2:l16}
            $\text{level}[u] = \text{level}[u] + 1$; \label{algo2:l17}
            
            \tcp{marking as OUT or UNDECIDED}
            
            $\text{mark}[u] = (level[u] == 0)$; \label{algo2:l22}
        }{
            \If{$state(u) == \text{OUT}$ \textbf{and} $\text{level}[u] == 0$}{ \label{algo2:l24}

                \tcp{marking as UNDECIDED}
                
                $\text{mark}[u] = \text{true}$; \label{algo2:l25}
            }
        }
    } \label{algo2:l28}

    syncMarkInsert(mark); \Comment{See Algorithm~\ref{alg:sync_mark_insert}}\label{algo2:l29}

    multiGpuMis($G_i$, mark, level); \Comment{Algorithm~\ref{algo1:luby_multi}}\label{algo2:l30}
}
\end{algorithm}
When a new edge (\textit{u}, \textit{v}) is inserted into the graph $G$, three cases arise based on the endpoint states. (i) $state(u) = \text{OUT}$ and $state(v) = \text{OUT}$: in this case neither independence nor maximality is affected. 
(ii) exactly one vertex has $state=\text{IN}$ and the other vertex has $state=\text{OUT}$: The vertex with $state=\text{OUT}$ gets a new vertex with $state=\text{IN}$, so its level increases by one. (iii) $state(u) = \text{IN}$ and $state(v) = \text{IN}$: 
independence is violated in this case. The lower priority vertex is revoked. Consequently, the \textit{level} of the neighbors of the revoked vertex is decremented by one. If any neighbor's level drops to zero, it transitions to $state(v) = \text{UNDECIDED}$.

Each undirected insertion appears as both directed pairs $(src, dst)$ and
$(dst, src)$, partitioned so that each GPU processes edges whose $src$ is
local to that GPU. The Algorithm~\ref{alg:insert} handles the update in three phases.

\paragraph{Phase 1: Conflict Identification} In the first phase
(Lines~\ref{algo2:l1}--\ref{algo2:p1end}), the algorithm processes all inserted edges (Line~\ref{algo2:l1}) and checks if $state(dst) = \text{IN}$
(Line~\ref{algo2:l2}). For each such edge, the GPU atomically increments $\text{level}[src]$
(Line~\ref{algo2:l3}). If both $state(src) = \text{IN}$ and $state(dst) = \text{IN}$, the independence property is
violated; the algorithm flags $src$ as revoked when $\pi(dst) > \pi(src)$
(Lines~\ref{algo2:l4}--\ref{algo2:l5}). The reciprocal edge $(dst, src)$ is evaluated by the GPU
owning $dst$, ensuring exactly one endpoint is revoked per conflict.

\paragraph{Phase 2: Neighbor Reduction} In the second phase (Lines~\ref{algo2:l7}--\ref{algo2:l13}),
for each revoked vertex $w$, the algorithm atomically decrements the level of all its
neighbors in $G'$ (which includes the newly inserted edges). Because these
neighbors may reside on remote GPUs, the decrements are accumulated locally
and synchronized via $\textsc{syncLevelInsert}$ (Line~\ref{algo2:l14}). After this phase, each vertex's
level reflects the net change in its count of neighbors with $state(v) = \text{IN}$.

\paragraph{Phase 3: State Resolution} In the final phase (Lines~\ref{algo2:l15}--\ref{algo2:l30}),
each GPU resolves the states of its local vertices; because each thread processes a distinct local vertex $u$, no atomic updates are required in this phase. A revoked vertex's
level was previously $\infty$ ($\text{UINT\_MAX}$). The increments from
Phase~1 and decrements from Phase~2 are offsets relative to this sentinel.
The correction on Line~\ref{algo2:l17} ($\text{level}[u] \gets \text{level}[u] + 1$)
exploits unsigned wraparound: adding 1 to $\text{UINT\_MAX}$ yields 0, so
the accumulated deltas now give the true count of remaining neighbors with $state(v) = \text{IN}$.
If this count is zero, the revoked vertex transitions to $state(v) = \text{UNDECIDED}$;
otherwise, it transitions to $state(v) = \text{OUT}$. This is achieved concisely by setting $\text{mark}[u]$ to true $(level[u] == 0)$ (Line~\ref{algo2:l22}). Additionally, any
non-revoked vertex with $state(v) = \text{OUT}$ whose level reaches zero, due to its neighbors with $state(v) = \text{IN}$
being revoked without gaining new ones, also transitions to $state(v) = \text{UNDECIDED}$
(Lines~\ref{algo2:l24}--\ref{algo2:l25}). Finally, all state changes are broadcast via $\textsc{syncMarkInsert}$
(Line~\ref{algo2:l29}) to maintain global consistency before the Luby algorithm is
re-invoked (Line~\ref{algo2:l30}).

\par Algorithm~\ref{alg:insert} is executed on the device as a series of kernel calls from the host. The above three phases correspond to separate GPU kernel launches. The host issues a \texttt{cudaDeviceSynchronize()} between Phase~1 and Phase~2 to ensure all conflict identification results are visible before neighbor reduction begins. The $\textsc{syncLevelInsert}$ (Line~\ref{algo2:l14}) and $\textsc{syncMarkInsert}$ (Line~\ref{algo2:l29}) calls ensure that the $\text{level}$ and $\text{mark}$ values are consistent across all devices. Finally, the host invokes $\textsc{multiGpuMis}$ (Line~\ref{algo2:l30}) to re-run the Luby algorithm on the remaining vertices with $state(v) = \text{UNDECIDED}$ to compute the new MIS.

\par Unlike $\textsc{syncLevel}$ in Algorithm~\ref{algo1:luby_multi}, which transmits \emph{increments}, $\textsc{syncLevelInsert}$ transmits the \emph{decrements} accumulated during neighbor-level reduction. The receiving GPU subtracts the received value from its local copy. $\textsc{syncMarkInsert}$ in Algorithm~\ref{alg:insert} is identical to $\textsc{syncMark}$ in Algorithm~\ref{algo1:luby_multi}.

\subsubsection{Decremental-MultiGPU-MIS} 
\REM{ \begin{algorithm}[t]
\caption{multiGpuEdgeDeletions}
\label{alg:delete}
\SetKwProg{Fn}{}{ \{}{\}}
\Fn(){\parbox[t]{0.85\linewidth}{multiGpuEdgeDeletions(\texttt{Graph} $G_i$, \texttt{bool *}mark, \\ \texttt{int *}level, $\Delta E^-$)}}{
    \ForEach{\parbox[t]{0.85\linewidth}{$(src, dst) \in \Delta E^- \mid state(src) == \text{OUT}$ \\ \textbf{and} $state(dst) == \text{IN}$ \textbf{in parallel}}}{
        $\text{level}[src] = \text{level}[src] - 1$;
    }

    \ForEach{local vertex $u$ with $state(u) == \text{OUT}$ \textbf{in parallel}}{
        \If{$\text{level}[u] == 0$}{
            $\text{mark}[u] = \text{UNDECIDED}$;

            Flag $u$ for broadcast;
        }
    }

    syncMark(mark);

    multiGpuMis($G_i$, mark, level);
}
\end{algorithm}
}
\begin{algorithm}[t]
\caption{Multi-GPU Decremental Updates}
\label{alg:delete}
\SetKwProg{Fn}{}{ \{}{\}}
\Fn(){{multiGpuEdgeDeletions(\textup{\texttt{Graph}} $G_i$, \textup{\texttt{Edges}} $\Delta E^-$,\\ \textup{\texttt{int *}} level, \textup{\texttt{bool *}} mark)}}{
    \ForEach{$(src, dst) \in \Delta E^-$ \textbf{in parallel}}{ \label{algo3:l1}
    \If{$state(src) == \text{OUT}$ \textbf{and} $state(dst) == \text{IN}$}{ \label{algo3:l2}
      \textbf{atomic}($\text{level}[src] = \text{level}[src] - 1$); \label{algo3:l3}
    }
    } \label{algo3:l5}
    
\tcp{cudaDeviceSynchronize()}
    
    \texttt{Sync\_With\_GPU()}; 
    
    \ForEach{local vertex $u$ \textbf{in parallel}}{ \label{algo3:l6}
        \If{$state(u) == \text{OUT}$ \textbf{and} $\text{level}[u] == 0$}{ \label{algo3:l7}

            \tcp{marking as UNDECIDED}
            
            $\text{mark}[u] = \text{true}$;\label{algo3:l8}

        }
    } \label{algo3:l10}

    syncMarkDelete(mark); \label{algo3:l11}
    
    multiGpuMis($G_i$, mark, level); \label{algo3:l12}
}
\end{algorithm}
Edge deletions cannot create independence violations, since no edge between
two vertices having $state$ value $\text{IN}$ exists in a valid MIS $I$. The two cases in deleting an edge (\textit{u}, \textit{v}) are:
(i) $state(u) = \text{OUT}$ and $state(v) = \text{OUT}$: This does not affect the MIS $I$. (ii) exactly one of the vertices has $state$=$\text{IN}$ and the other has $state$= $\text{OUT}$: 
 the vertex having $state$=$\text{OUT}$ loses a neighbor having $state$=$\text{IN}$, decreasing
its \textit{level} by 1. If its \textit{level} drops to zero, the vertex no longer has any neighbor having $state$=$\text{IN}$, hence the maximality property is violated, causing it to transition to $state$ $\text{UNDECIDED}$.

As with insertions, each undirected deletion appears as both directed pairs, partitioned by \textit{src} vertex ownership. The Algorithm~\ref{alg:delete} handles this in two phases.

\paragraph{Phase 1: Level Reduction} The first phase (Lines~\ref{algo3:l1}--\ref{algo3:l5})
processes each deleted edge where $state(src) = \text{OUT}$ and $state(dst) = \text{IN}$ (Line~\ref{algo3:l2}),
atomically decrementing $\text{level}[src]$ (Line~\ref{algo3:l3}).\REM{ Edges between two vertices with $state(v) = \text{OUT}$,
or those evaluated from the perspective of the endpoint with $state(v) = \text{IN}$, do not affect any
level and are skipped.}

\paragraph{Phase 2: State Resolution} The second phase (Lines~\ref{algo3:l6}--\ref{algo3:l10})
evaluates all local vertices with $state(v) = \text{OUT}$ in parallel (Line~\ref{algo3:l6}). If an OUT vertex's
level drops to zero (Line~\ref{algo3:l7}), it is no longer adjacent to any vertex with $state(v) = \text{IN}$ in
$G'$, so it transitions to $state(v) = \text{UNDECIDED}$ by setting its \textit{mark} value to \texttt{true} (Line~\ref{algo3:l8}). These state changes are synchronized across all the GPUs via $\textsc{syncMarkDelete}$ (Line~\ref{algo3:l11}). 
Note that deletions do not require $\textsc{syncLevel}$, since each GPU only decrements levels of its own local vertices. Finally, the Luby algorithm is re-invoked (Line~\ref{algo3:l12}) on the vertices having $state(v) = \text{UNDECIDED}$ to re-compute the global MIS.

\par Algorithm~\ref{alg:delete} follows a simpler execution compared to insertions. The two phases are executed as sequential GPU kernel launches. Since each GPU only decrements levels of its own local vertices, no $\textsc{syncLevel}$ call is required between phases. The $\textsc{syncMarkDelete}$ (Line~\ref{algo3:l11}) call ensures the $\text{mark}$ is consistent across all devices. The host then invokes $\textsc{multiGpuMis}$ (Line~\ref{algo3:l12}) to resolve the remaining vertices with $state(v) = \text{UNDECIDED}$ to compute the new MIS.

\par Unlike $\textsc{syncMark}$ in Algorithm ~\ref{algo1:luby_multi}, which broadcasts \emph{OUT} transitions ($\text{mark} = \texttt{false}$), $\textsc{syncMarkDelete}$ broadcasts \emph{UNDECIDED} transitions ($\text{mark} = \texttt{true}$, $\text{level} = 0$) to all GPUs, since deletion can only cause vertices to lose IN neighbours and become undecided.

\subsection{Working Example}

\subsubsection{Static MIS Computation} We demonstrate using the graph in Figure~\ref{fig:DMG-MIS_luby}, partitioned across GPU~0 ($\{A,\allowbreak B,\allowbreak C,\allowbreak D,\allowbreak E\}$) and GPU~1 ($\{F,\allowbreak G,\allowbreak H,\allowbreak I,\allowbreak J\}$), with cross-GPU edges $(C,F)$ and $(C,I)$. All vertices start as UNDECIDED.

In Iteration~1, \texttt{MISLocal\_gpu} identifies $\{A,F,G\}$ as having the highest priorities among their UNDECIDED neighbors, transitioning them to IN. The $\textsc{setToOut}$ kernel then transitions their neighbors $\{B,C,H\}$ to OUT, incrementing levels: $B$ to~1 (from~$A$), $C$ to~1 (from remote neighbor~$F$, accumulated on GPU~1 and transmitted via $\textsc{syncLevel}$), and $H$ to~2 (from~$F$, $G$). $\textsc{syncMark}$ broadcasts all state changes to ensure global consistency. Vertices $\{D,E,I,J\}$ remain UNDECIDED.

In Iteration~2, $\{D,I,J\}$ are the highest-priority vertices among their UNDECIDED neighbors and transitions to IN. $\textsc{setToOut}$ increments $E$'s level to~1 (from~$D$), making it OUT, and increments $C$'s level locally (from~$D$) and remotely (from~$I$ on GPU~1, synchronized via $\textsc{syncLevel}$), updating $C$'s level to~3 and $H$ level increments to~3. After $\textsc{syncMark}$ broadcasts these changes, no UNDECIDED vertices remain and the algorithm terminates, yielding the final MIS $\{A,D,F,G,I,J\}$.

\begin{figure}[htbp] \centering
\begin{tikzpicture}[ scale=0.55, every node/.style={font=\scriptsize}, 
    node/.style={ circle, draw=black, very thick, minimum size=0.4cm, fill=white, inner sep=0.5pt }, 
    active/.style={ circle, draw=black, very thick, fill=blue!80!black, text=white, minimum size=0.4cm, inner sep=0.5pt },
    outnode/.style={ circle, draw=black, very thick, fill=gray!20, minimum size=0.4cm, inner sep=0.5pt },
    edge/.style={black, thick}, 
    split/.style={red, dashed, very thick} ]

    \node[node, draw=none] (GPU0) at (1, 3.5) {GPU0}; 
    \node[node, draw=none] (GPU1) at (5, 3.5) {GPU1};
    
    \node[active] (A) at (0,2) {$A:9$};
    \node[outnode] (B) at (2,2) {$B:2$}; 
    \node[outnode] (C) at (2,0) {$C:7$}; 
    \node[active] (D) at (2,-2) {$D:3$}; 
    \node[outnode] (E) at (0,-2) {$E:1$};
    
    \node[active] (F) at (4,0) {$F:8$};
    \node[active] (I) at (4,-2) {$I:5$};
    \node[outnode] (H) at (6,0) {$H:4$};
    \node[active] (G) at (6,-2) {$G:6$};
    \node[active] (J) at (6,2) {$J:0$};

    \draw[edge] (A)--(B); \draw[edge] (B)--(C);
    \draw[edge] (C)--(D); \draw[edge] (D)--(E); 
    \draw[edge] (C)--(F); \draw[edge] (C)--(I); 
    \draw[edge] (H)--(F); \draw[edge] (H)--(G); \draw[edge] (H)--(J);

    \draw[split] (3,-3.0)--(3,4.0); 
\end{tikzpicture}
\caption{Initial graph partitioned across two GPUs. \REM{Each vertex label shows its ID and priority $\pi$ (e.g., $A{:}9$ means $\pi(A)=9$). The dashed red line marks the GPU partition boundary. Colored (blue) vertices are IN the MIS; gray vertices are OUT.}}
\Description{Diagram of a 10-vertex graph partitioned across GPU0 (vertices A through E, with A and D in the MIS) and GPU1 (vertices F through J, with F, G, I, and J in the MIS), with cross-GPU edges (C,F) and (C,I).}
\label{fig:DMG-MIS_luby} \end{figure}

\subsubsection{Dynamic Incremental Updates} 
\begin{figure}
\centering
\subfloat[Insertion:\{(C,E),(C,G),(E,D)\}\label{figinsdel:ins}]{%
\begin{minipage}{0.48\columnwidth}
\centering
\resizebox{\linewidth}{!}{
\begin{tikzpicture}[ scale=0.55, every node/.style={font=\scriptsize}, 
    node/.style={ circle, draw=black, very thick, minimum size=0.4cm, fill=white, inner sep=0.5pt }, 
    active/.style={ circle, draw=black, very thick, fill=blue!80!black, text=white, minimum size=0.4cm, inner sep=0.5pt },
    outnode/.style={ circle, draw=black, very thick, fill=gray!20, minimum size=0.4cm, inner sep=0.5pt },
    edge/.style={black, thick}, 
    newedge/.style={magenta, ultra thick},
    split/.style={red, dashed, very thick} ]

    \node[node, draw=none] (GPU0) at (1, 3.5) {GPU0}; 
    \node[node, draw=none] (GPU1) at (5, 3.5) {GPU1};
    
    \node[active] (A) at (0,2) {$A:9$};
    \node[outnode] (B) at (2,2) {$B:2$}; 
    \node[outnode] (C) at (2,0) {$C:7$}; 
    \node[outnode] (D) at (2,-2) {$D:3$}; 
    \node[active] (E) at (0,-2) {$E:1$};
    
    \node[active] (F) at (4,0) {$F:8$};
    \node[active] (I) at (4,-2) {$I:5$};
    \node[outnode] (H) at (6,0) {$H:4$};
    \node[active] (G) at (6,-2) {$G:6$};
    \node[active] (J) at (6,2) {$J:0$};

    \draw[edge] (A)--(B); \draw[edge] (B)--(C);
    \draw[edge] (C)--(D); \draw[edge] (D)--(E); 
    \draw[edge] (C)--(F); \draw[edge] (C)--(I); 
    \draw[edge] (H)--(F); \draw[edge] (H)--(G); \draw[edge] (H)--(J);

    \draw[newedge] (C) -- (G);
    \draw[newedge] (C) -- (E);
    \draw[newedge] (D) -- (I);

    \draw[split] (3,-3.0)--(3,4.0); 
\end{tikzpicture}
}
\end{minipage}
}
\hfill
\subfloat[Deletion:\{(B,C),(C,I)(D,E)\}\label{figinsdel:del}]{%
\begin{minipage}{0.48\columnwidth}
\centering
\resizebox{\linewidth}{!}{
\begin{tikzpicture}[ scale=0.55, every node/.style={font=\scriptsize}, 
    node/.style={ circle, draw=black, very thick, minimum size=0.4cm, fill=white, inner sep=0.5pt }, 
    active/.style={ circle, draw=black, very thick, fill=blue!80!black, text=white, minimum size=0.4cm, inner sep=0.5pt },
    outnode/.style={ circle, draw=black, very thick, fill=gray!20, minimum size=0.4cm, inner sep=0.5pt },
    edge/.style={black, thick}, 
    deledge/.style={red!80!black, ultra thick, dashed},
    split/.style={red, dashed, very thick} ]

    \node[node, draw=none] (GPU0) at (1, 3.5) {GPU0}; 
    \node[node, draw=none] (GPU1) at (5, 3.5) {GPU1};
    
    \node[active] (A) at (0,2) {$A:9$};
    \node[outnode] (B) at (2,2) {$B:2$}; 
    \node[outnode] (C) at (2,0) {$C:7$}; 
    \node[active] (D) at (2,-2) {$D:3$}; 
    \node[active] (E) at (0,-2) {$E:1$};
    
    \node[active] (F) at (4,0) {$F:8$};
    \node[active] (I) at (4,-2) {$I:5$};
    \node[outnode] (H) at (6,0) {$H:4$};
    \node[active] (G) at (6,-2) {$G:6$};
    \node[active] (J) at (6,2) {$J:0$};

    \draw[edge] (A)--(B); 
    \draw[edge] (C)--(D); 
    \draw[edge] (C)--(F); 
    \draw[edge] (H)--(F); \draw[edge] (H)--(G); \draw[edge] (H)--(J);

    \draw[split] (3,-3.0)--(3,4.0); 
\end{tikzpicture}
}
\end{minipage}
}
\caption{MIS after batch updates on Graph in Figure~\ref{fig:DMG-MIS_luby}}
\Description{Two-panel diagram showing the updated MIS on the partitioned graph after a batch of edge insertions on the left and a batch of edge deletions on the right.}
\label{figinsdel}
\end{figure}
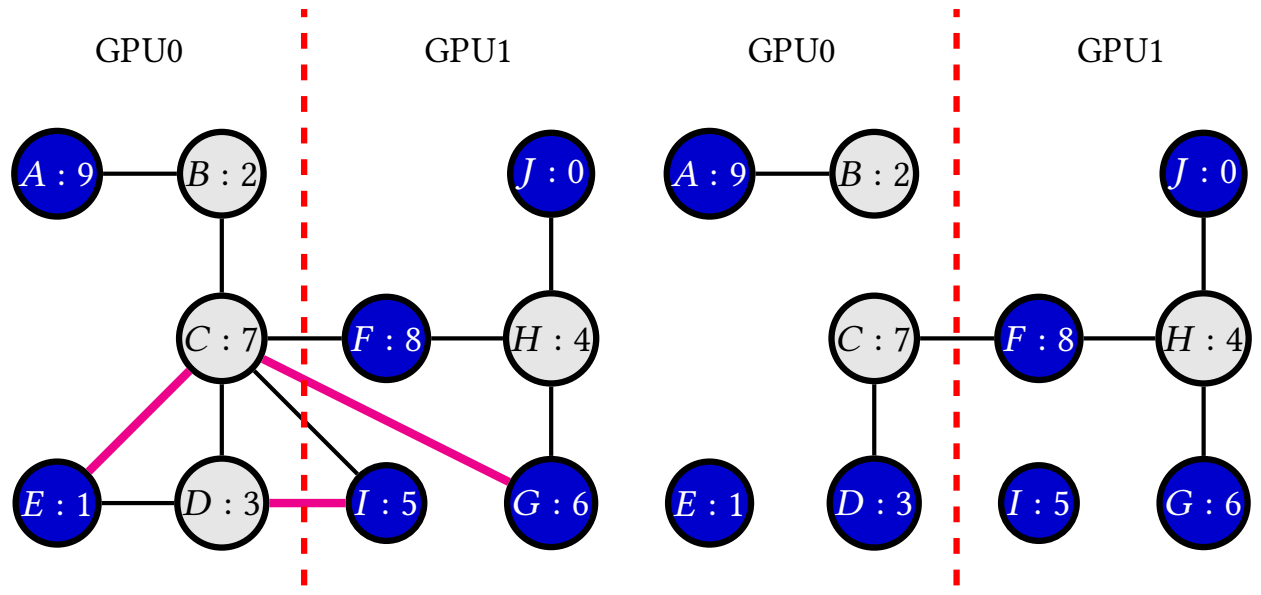

The Figure~\ref{figinsdel:ins}
illustrates the graph after a batch of edge insertions on graph in Figure ~\ref{fig:DMG-MIS_luby}: $\Delta E^+ = \{ (C,G), (C,E), (D,I) \}$. Here the edges are inserted in both directions. For example consider the edge $(C,G)$ being inserted: when $(C,G)$ is inserted in GPU0, the edge $(G,C)$ gets inserted in GPU1 simultaneously.
\REM{
\begin{figure}[htbp]
  \centering
  \begin{minipage}{0.48\textwidth}
     \centering
 \caption{Insertion of a batch of edges} \label{fig:DMG-MIS_insert}
   \end{minipage}\hfill
\end{figure}

\begin{figure*}[htbp]
    \centering
    \begin{minipage}{0.48\textwidth}
        \centering
        \begin{minipage}{0.48\columnwidth}
\centering
\begin{tikzpicture}[ scale=0.55, every node/.style={font=\scriptsize}, 
    node/.style={ circle, draw=black, very thick, minimum size=0.4cm, fill=white, inner sep=0.5pt }, 
    active/.style={ circle, draw=black, very thick, fill=blue!80!black, text=white, minimum size=0.4cm, inner sep=0.5pt },
    outnode/.style={ circle, draw=black, very thick, fill=gray!20, minimum size=0.4cm, inner sep=0.5pt },
    edge/.style={black, thick}, 
    newedge/.style={magenta, ultra thick, dashed},
    split/.style={red, dashed, very thick} ]

    \node[node, draw=none] (GPU0) at (1, 3.5) {GPU0}; 
    \node[node, draw=none] (GPU1) at (5, 3.5) {GPU1};
    
    \node[active] (A) at (0,2) {$A:9$};
    \node[outnode] (B) at (2,2) {$B:8$}; 
    \node[outnode] (C) at (2,0) {$C:7$}; 
    \node[active] (D) at (2,-2) {$D:3$}; 
    \node[outnode] (E) at (0,-2) {$E:1$};
    
    \node[active] (F) at (4,0) {$F:8$};
    \node[active] (I) at (4,-2) {$I:5$};
    \node[outnode] (H) at (6,0) {$H:4$};
    \node[active] (G) at (6,-2) {$G:6$};
    \node[active] (J) at (6,2) {$J:7$};

    \draw[edge] (A)--(B); \draw[edge] (B)--(C);
    \draw[edge] (C)--(D); \draw[edge] (D)--(E); 
    \draw[edge] (C)--(F); \draw[edge] (C)--(I); 
    \draw[edge] (H)--(F); \draw[edge] (H)--(G); \draw[edge] (H)--(J);

    \draw[newedge] (C) -- (G);
    \draw[newedge] (C) -- (E);
    \draw[newedge] (D) -- (I);

    \draw[split] (3,-3.0)--(3,4.0); 
\end{tikzpicture}
\subcaption{Before Insertion (incoming edges dashed)}
\end{minipage}\hfill
\begin{minipage}{0.48\columnwidth}
\centering
\begin{tikzpicture}[ scale=0.55, every node/.style={font=\scriptsize}, 
    node/.style={ circle, draw=black, very thick, minimum size=0.4cm, fill=white, inner sep=0.5pt }, 
    active/.style={ circle, draw=black, very thick, fill=blue!80!black, text=white, minimum size=0.4cm, inner sep=0.5pt },
    outnode/.style={ circle, draw=black, very thick, fill=gray!20, minimum size=0.4cm, inner sep=0.5pt },
    edge/.style={black, thick}, 
    newedge/.style={magenta, ultra thick},
    split/.style={red, dashed, very thick} ]

    \node[node, draw=none] (GPU0) at (1, 3.5) {GPU0}; 
    \node[node, draw=none] (GPU1) at (5, 3.5) {GPU1};
    
    \node[active] (A) at (0,2) {$A:9$};
    \node[outnode] (B) at (2,2) {$B:8$}; 
    \node[outnode] (C) at (2,0) {$C:7$}; 
    \node[outnode] (D) at (2,-2) {$D:3$}; 
    \node[active] (E) at (0,-2) {$E:1$};
    
    \node[active] (F) at (4,0) {$F:8$};
    \node[active] (I) at (4,-2) {$I:5$};
    \node[outnode] (H) at (6,0) {$H:4$};
    \node[active] (G) at (6,-2) {$G:6$};
    \node[active] (J) at (6,2) {$J:7$};

    \draw[edge] (A)--(B); \draw[edge] (B)--(C);
    \draw[edge] (C)--(D); \draw[edge] (D)--(E); 
    \draw[edge] (C)--(F); \draw[edge] (C)--(I); 
    \draw[edge] (H)--(F); \draw[edge] (H)--(G); \draw[edge] (H)--(J);

    \draw[newedge] (C) -- (G);
    \draw[newedge] (C) -- (E);
    \draw[newedge] (D) -- (I);

    \draw[split] (3,-3.0)--(3,4.0); 
\end{tikzpicture}
\subcaption{After Insertion (recomputed MIS)}
\end{minipage}
        \caption{Insertion of a batch of edges} \label{fig:DMG-MIS_insert}
    \end{minipage}\hfill
    \begin{minipage}{0.48\textwidth}
        \centering
        \begin{minipage}{0.48\columnwidth}
\centering
\begin{tikzpicture}[ scale=0.55, every node/.style={font=\scriptsize}, 
    node/.style={ circle, draw=black, very thick, minimum size=0.4cm, fill=white, inner sep=0.5pt }, 
    active/.style={ circle, draw=black, very thick, fill=blue!80!black, text=white, minimum size=0.4cm, inner sep=0.5pt },
    outnode/.style={ circle, draw=black, very thick, fill=gray!20, minimum size=0.4cm, inner sep=0.5pt },
    edge/.style={black, thick}, 
    deledge/.style={red!80!black, ultra thick, dashed},
    split/.style={red, dashed, very thick} ]

    \node[node, draw=none] (GPU0) at (1, 3.5) {GPU0}; 
    \node[node, draw=none] (GPU1) at (5, 3.5) {GPU1};
    
    \node[active] (A) at (0,2) {$A:9$};
    \node[outnode] (B) at (2,2) {$B:8$}; 
    \node[outnode] (C) at (2,0) {$C:7$}; 
    \node[active] (D) at (2,-2) {$D:3$}; 
    \node[outnode] (E) at (0,-2) {$E:1$};
    
    \node[active] (F) at (4,0) {$F:8$};
    \node[active] (I) at (4,-2) {$I:5$};
    \node[outnode] (H) at (6,0) {$H:4$};
    \node[active] (G) at (6,-2) {$G:6$};
    \node[active] (J) at (6,2) {$J:7$};

    \draw[edge] (A)--(B); 
    \draw[deledge] (B)--(C);
    \draw[edge] (C)--(D); 
    \draw[deledge] (D)--(E); 
    \draw[edge] (C)--(F); 
    \draw[deledge] (C)--(I); 
    \draw[edge] (H)--(F); \draw[edge] (H)--(G); \draw[edge] (H)--(J);

    \draw[split] (3,-3.0)--(3,4.0); 
\end{tikzpicture}
\subcaption{Before Deletion (outgoing edges dashed)}
\end{minipage}\hfill
\begin{minipage}{0.48\columnwidth}
\centering
\begin{tikzpicture}[ scale=0.55, every node/.style={font=\scriptsize}, 
    node/.style={ circle, draw=black, very thick, minimum size=0.4cm, fill=white, inner sep=0.5pt }, 
    active/.style={ circle, draw=black, very thick, fill=blue!80!black, text=white, minimum size=0.4cm, inner sep=0.5pt },
    outnode/.style={ circle, draw=black, very thick, fill=gray!20, minimum size=0.4cm, inner sep=0.5pt },
    edge/.style={black, thick}, 
    deledge/.style={red!80!black, ultra thick, dashed},
    split/.style={red, dashed, very thick} ]

    \node[node, draw=none] (GPU0) at (1, 3.5) {GPU0}; 
    \node[node, draw=none] (GPU1) at (5, 3.5) {GPU1};
    
    \node[active] (A) at (0,2) {$A:9$};
    \node[outnode] (B) at (2,2) {$B:8$}; 
    \node[outnode] (C) at (2,0) {$C:7$}; 
    \node[active] (D) at (2,-2) {$D:3$}; 
    \node[active] (E) at (0,-2) {$E:1$};
    
    \node[active] (F) at (4,0) {$F:8$};
    \node[active] (I) at (4,-2) {$I:5$};
    \node[outnode] (H) at (6,0) {$H:4$};
    \node[active] (G) at (6,-2) {$G:6$};
    \node[active] (J) at (6,2) {$J:7$};

    \draw[edge] (A)--(B); 
    \draw[edge] (C)--(D); 
    \draw[edge] (C)--(F); 
    \draw[edge] (H)--(F); \draw[edge] (H)--(G); \draw[edge] (H)--(J);

    \draw[split] (3,-3.0)--(3,4.0); 
\end{tikzpicture}
\subcaption{After Deletion (recomputed MIS)}
\end{minipage}
        \caption{Deletion of a batch of edges} \label{fig:DMG-MIS_delete}
    \end{minipage}
\end{figure*}
}
\REM{In the first phase, the algorithm processes each inserted edge where
the destination is IN. For edge $(C,G)$, $G$ is IN, so
$\text{level}[C]$ is incremented by 1. Because $C$ is OUT, no IN-IN
conflict arises. For edge $(C,E)$, both $C$ and $E$ are OUT, so neither
independence nor maximality is affected and no state changes occur. For
edge $(D,I)$, $I$ is IN, so $\text{level}[D]$ is incremented by 1.
Because both $D$ and $I$ are IN, and $\pi(I) = 5 > \pi(D) = 3$, $D$ is
flagged as revoked.

In the second phase, the algorithm decrements the $\text{level}$ of all
neighbors of the revoked vertex $D$ in the updated graph $G'$. $D$'s
neighbors in $G'$ are $\{C, E, I\}$ (its original neighbors $C$ and
$E$, plus the new neighbor $I$). The $\text{level}$ of $C$, $E$, and
$I$ are each decremented by 1. Because $I$ resides on GPU~1, GPU~0
accumulates this decrement locally and synchronizes it via
$\textsc{syncLevel}$.

In the final phase, the revoked vertex $D$'s $\text{level}$ is
corrected by incrementing it by 1, yielding a $\text{level}$ of 1
(reflecting its one valid IN neighbor, $I$). Since its $\text{level}
> 0$, $D$ transitions to the OUT state. Vertex $C$'s $\text{level}$ was
initially 3 (from IN neighbors $D$, $F$, and $I$); it increased by 1 in
the first phase (due to the new edge to $G$), and then decreased by 1
in the second phase (due to $D$'s revocation), resulting in a net
$\text{level}$ of 3, keeping $C$ in the OUT state. Vertex $E$'s
$\text{level}$ was 1 (from $D$), but decreased by 1 in the second
phase, resulting in a $\text{level}$ of 0. Consequently, $E$
transitions to the UNDECIDED state.

The newly UNDECIDED vertex $E$ is broadcast via $\textsc{syncMarkInsert}$. The
multi-GPU Luby algorithm then resumes specifically for the UNDECIDED
vertices. $E$'s neighbors in $G'$ are $D$ (OUT) and $C$ (OUT, from the
new edge), so $E$ has no UNDECIDED neighbors with a higher priority and
immediately transitions to IN. The final updated MIS is $\{A, E, F, G,
I, J\}$.}
Algorithm~\ref{alg:insert} executes in three phases. In Phase~1 (Lines~\ref{algo2:l1}--\ref{algo2:l6}), for each inserted edge whose destination is IN, $\text{level}[src]$ is incremented. Edge $(C,G)$: $G$ is IN, so $\text{level}[C]$ increases to~4; $C$ is OUT, so no conflict. Edge $(C,E)$: both are OUT, so no action. Edge $(D,I)$: $I$ and $D$ are IN, so $\text{level}[D]$ and $\text{level}[I]$ are incremented; and $\pi(I) > \pi(D)$, so $D$ is flagged as revoked.

In Phase~2 (Lines~\ref{algo2:l7}--\ref{algo2:l13}), the levels of all neighbors of revoked vertex $D$ in $G'$ are decremented. $D$'s neighbors in $G'$ are $\{C, E, I\}$. Because $I$ resides on GPU~1, GPU~0 accumulates this decrement locally and synchronizes it via $\textsc{syncLevelInsert}$ (Line~\ref{algo2:l14}).

In Phase~3 (Lines~\ref{algo2:l15}--\ref{algo2:l28}), $D$'s level is corrected (Line~\ref{algo2:l17}): the increment yields a level of~1 (reflecting its one remaining IN neighbor, $I$), so $D$ transitions to OUT. $C$'s level was~3, increased by~1 in Phase~1 (edge to $G$), then decreased by~1 in Phase~2 ($D$'s revocation), netting~3, so $C$ remains OUT. $E$'s level was~1 (from $D$), decreased by~1 in Phase~2, yielding~0, so $E$ transitions to UNDECIDED. These state changes are broadcast via $\textsc{syncMarkInsert}$ (Line~\ref{algo2:l29}). The Luby algorithm resumes (Line~\ref{algo2:l30}): $E$'s neighbors in $G'$ are $D$ (OUT) and $C$ (OUT), so $E$ transitions to IN, yielding the final MIS $\{A, E, F, G, I, J\}$ (See Figure~\ref{figinsdel:ins}).
\subsubsection{Dynamic Decremental Updates} Figure~\ref{figinsdel:del}
illustrates the response to a batch of edge deletions from the initial
graph (Figure~\ref{fig:DMG-MIS_luby}): $\Delta E^- = \{ (B, C), (C, I), (D, E) \}$. 
\REM{The algorithm processes each deleted edge where the source is OUT and
the destination is IN. For edge $(B, C)$, both $B$ and $C$ are OUT, so
no $\text{level}$ change is required. For edge $(C, I)$, $C$ is OUT and
$I$ is IN, so $\text{level}[C]$ is decremented by 1. $C$'s original
IN neighbors were $D$, $F$, and $I$, giving it an initial
$\text{level}$ of 3. After the decrement, $C$'s $\text{level}$ becomes
2 (as it remains covered by $D$ and $F$), so $C$ remains OUT. For edge
$(D, E)$, $E$ is OUT and $D$ is IN, so $\text{level}[E]$ is
decremented by 1. $E$'s $\text{level}$ was 1 (from its only IN
neighbor, $D$). After the decrement, $\text{level}[E] = 0$, so $E$
transitions to the UNDECIDED state.
The newly UNDECIDED vertex $E$ is broadcast via $\textsc{syncMarkDelete}$ to all other GPUs. The Luby algorithm is then re-run on the UNDECIDED
vertices. $E$ now has no neighbors in the updated graph (its only edge
$(D,E)$ was deleted), meaning it has no UNDECIDED or IN neighbors.
Thus, it transitions to IN. The final updated MIS is $\{A, D, E, F, G,
I, J\}$.}
Algorithm~\ref{alg:delete} processes only edges where $state(src)=\text{OUT}$ and $state(dst)=\text{IN}$. Edge $(B,C)$: both are OUT, so it is skipped. Edge $(C,I)$: $C$ is OUT and $I$ is IN, so $\text{level}[C]$ is decremented from~3 to~2; $C$ remains covered by IN neighbors $D$ and $F$, so it stays OUT. Edge $(D,E)$: from the directed pair $(E,D)$ where $E$ is OUT and $D$ is IN, $\text{level}[E]$ is decremented from~1 to~0; $E$ has lost its only IN neighbor and transitions to UNDECIDED. This transition is broadcast to all GPUs via $\textsc{syncMarkDelete}$ (Line~\ref{algo3:l11}). The Luby algorithm resumes (Line~\ref{algo3:l12}): $E$ now has no neighbors in $G'$ (its only edge was deleted), so it immediately transitions to IN, yielding the final MIS $\{A, D, E, F, G, I, J\}$.

\section{Experimental Evaluation}\label{sec:expt}
\subsection{Experimental Setup}
 \textbf{Hardware:} Experiments use NVIDIA RTX 4090 GPUs 
 (24 GB GDDR6X) connected via PCIe 4.0, staging inter-GPU transfers through pinned host memory. All GPUs reside on the same NUMA node.

\textbf{Software:} We use CUDA Toolkit 13.2.2. Each GPU is managed by a separate host thread (one process, multiple GPUs). Inter-GPU communication is performed via peer-to-peer memory copies through PCIe, without MPI.

\textbf{Datasets:} We evaluate 9 graphs of varying sizes and structures (Table~\ref{tab:datasets}), including brain networks~\cite{25}, a social network~\cite{26}, an Erdős–Rényi graph~\cite{27}, a road network~\cite{24}, and three RMAT~\cite{28} synthetic graphs.

			\begin{table}[htbp] \centering \resizebox{\columnwidth}{!}{%
			\begin{tabular}{llrrrr} \toprule
				\textbf{Abbr.} & \textbf{Dataset} & \textbf{$|V|$} &
				\textbf{$|E|$} & \textbf{$d_{avg}$} & \textbf{$d_{max}$} \\ \midrule 
				BN1 & brain networks & 1.8M & 202M & 112 & 758 \\ 
				BN2 & brain networks & 1.8M & 67M & 37 & 1,539 \\
				CO & com-orkut & 3.1M & 117M & 38 & 33,313 \\ 
				ER & Erdős–Rényi & 2.0M & 150M & 75 & 215 \\ 
				USA & roadnetwork-usa & 23.9M & 28.8M & 1 & 9 \\
				CF & com-friendster & 65.6M & 1.8B & 27 & 5,214 \\ 
				RMAT1 & RMAT & 67.1M & 268.4M & 4 & 16,343 \\ 
				RMAT2 & RMAT & 67.1M & 536.8M & 8 & 32,477 \\ 
				RMAT3 & RMAT & 67.1M & 1.07B & 16 & 65,102 \\
			\bottomrule \end{tabular}} \caption{Graph characteristics.} \label{tab:datasets} \end{table}
	\subsubsection{Methodology} We evaluate two setups: (1) \textbf{Static}: our static MIS vs. MG-MIS~\cite{18}; and (2) \textbf{Dynamic}: our batched dynamic update algorithm. Static evaluations use CSR arrays, while dynamic updates use faimGraph. In the static setup, we only require $mark$ to compute the MIS; the $level$ array is used exclusively in the dynamic algorithms to track the independent cover size for efficient incremental updates.  Dynamic execution times for insertions and deletions are measured across five batch sizes (1K, 10K, 0.1\%, 1\%, and 10\% of $|E|$) and reported as speedups over full static recomputation.

	\textbf{Batch Generation.} Edge update batches are generated using utility provided by faimGraph~\cite{19}. For each batch size, the source and destination vertices of each edge are selected
	independently and uniformly at random from the vertex set. For insertions, only edges not already present in the graph are inserted. For deletions, candidate edges are generated
	using the same random procedure, but only those that correspond to existing
	edges in the graph are retained.

	\subsection{Static Performance}

	Table~\ref{tab:static_perf} compares the static MIS performance of DMG-MIS
	against MG-MIS across all graphs. We report per-graph speedups and
	the geometric mean. Entries where MG-MIS runs out of memory are marked OOM
	and excluded from the mean.

	\textbf{Single-GPU.} On a single GPU, DMG-MIS achieves a geometric mean
	speedup of $\sim$1.37$\times$ over MG-MIS. Note that MG-MIS faces OOM on the \textit{com-friendster} (CF) 
	graph because it uses 64-bit vertex indices. DMG-MIS instead uses 32-bit vertex indices, which is 
	what the faimGraph uses.
	The improvement comes from implementation-level changes, which includes warp-cooperative neighbor
	traversal and a reduction in the number of kernel
	launches. Because adjacency lists are stored contiguously in memory, having all 32 threads in a warp collaboratively traverse the neighbors of one active vertex at a time yields coalesced global memory reads and balances work across skewed vertex degrees, avoiding the lane underutilization and warp divergence of thread-per-vertex execution. By algorithm design, DMG-MIS also requires fewer kernel launches per
	Luby round compared to MG-MIS, minimizing launch overhead.
\begin{table}[htbp]
\centering
\footnotesize
\setlength{\tabcolsep}{4pt}
\begin{tabular}{lrrrrrr}
\toprule
 & \multicolumn{3}{c}{\textbf{DMG-MIS}} & \multicolumn{3}{c}{\textbf{MG-MIS}} \\ \cmidrule(lr){2-4} \cmidrule(l){5-7}
\textbf{Graph} & \textbf{1 GPU} & \textbf{2 GPUs} & \textbf{4 GPUs} & \textbf{1 GPU} & \textbf{2 GPUs} & \textbf{4 GPUs} \\ \midrule
BN1 & 10 & 10 & 10 & 3 & 744 & 911 \\
BN2 & 8 & 8 & 8 & 3 & 174 & 348 \\
CO & 8 & 9 & 9 & 2 & 193 & 213 \\
ER & 10 & 11 & 11 & 3 & 175 & 247 \\
USA & 6 & 6 & 6 & 2 & 5 & 7 \\
CF & 8 & 9 & 9 & OOM & OOM & 342 \\
RMAT1 & 7 & 7 & 7 & 2 & 35 & 39 \\
RMAT2 & 8 & 8 & 8 & 2 & 43 & 51 \\
RMAT3 & 8 & 9 & 9 & 2 & 58 & 73 \\
\bottomrule
\end{tabular}
\caption{Iteration counts for MG-MIS and DMG-MIS.}
\label{tab:iterations}
\end{table} 

	\textbf{Multi-GPU.} On 2 and 4 GPUs, the geometric mean speedups over MG-MIS
	are $19.10\times$ and $16.91\times$ respectively. The source of this improvement is algorithmic. In DMG-MIS, when a vertex $v$'s state transitions to $state(v) = \text{IN}$, it sends this information to all of its remote neighbors in a single communication round, and the receiving GPU applies the update to all affected vertices. This resolves
	states in significantly fewer rounds than MG-MIS. Table~\ref{tab:iterations} reports
	the iteration counts across all graphs. Each Luby iteration in MG-MIS executes the local 
	computation phase twice, synchronizes, and then executes two further local rounds; DMG-MIS, in each iteration, executes only 
	a single local round before synchronizing. On a single GPU, where no inter-GPU synchronization is needed, this lets MG-MIS 
	converge in fewer total iterations than DMG-MIS. As the number of GPU partitions grows, the iteration count for both 
	algorithms either increases or stays the same, since partitioning introduces cross-device dependencies.

MG-MIS partitions vertices equally across GPUs, i.e., $|V|/N$ per GPU where
	$N$ is number of GPUs. DMG-MIS instead partitions the graph by assigning
	each GPU an approximately equal share of $|V| + |E|$. This prevents GPUs
	with relatively denser partitions from becoming bottlenecks and ensures a more balanced
	memory distribution across devices. This allows DMG-MIS to achieve significant
	performance gains on graphs with a relatively higher number of edges. For instance, on BN1, DMG-MIS is
	$259.30\times$ faster than MG-MIS on 2 GPUs, and $210.90\times$ faster on
	4 GPUs, avoiding the severe load imbalance that hinders MG-MIS.

\begin{table}[htbp]
\centering
\begin{tabular}{lrrr}
\toprule
\textbf{Graph} & \textbf{1 GPU} & \textbf{2 GPUs} & \textbf{4 GPUs} \\ \midrule
BN1 & 2.59$\times$ & 259.30$\times$ & 210.90$\times$ \\
BN2 & 2.08$\times$ & 47.50$\times$ & 63.57$\times$ \\
CO & 1.63$\times$ & 33.80$\times$ & 31.06$\times$ \\
ER & 1.38$\times$ & 23.92$\times$ & 30.34$\times$ \\
USA & 1.18$\times$ & 1.60$\times$ & 0.96$\times$ \\
CF & - & - & 15.51$\times$ \\
RMAT1 & 0.92$\times$ & 9.20$\times$ & 7.34$\times$ \\
RMAT2 & 1.04$\times$ & 10.33$\times$ & 8.44$\times$ \\
RMAT3 & 0.93$\times$ & 11.69$\times$ & 9.69$\times$ \\ \midrule
\textbf{Mean} & 1.37$\times$ & 19.10$\times$ & 16.91$\times$ \\
\bottomrule
\end{tabular}
\caption{Static MIS speedup: DMG-MIS over MG-MIS.}
\label{tab:static_perf}
\end{table}
	
\textbf{Scaling behavior.} DMG-MIS's absolute runtime increases from 1 to 2 to 4 GPUs on every graph. Each Luby iteration requires a full synchronization round in which every GPU broadcasts its state changes to all other GPUs via PCIe. Unlike MG-MIS, which polls remote neighbors on demand, DMG-MIS publishes state changes eagerly, making communication cost high. Thus, the multi-GPU setting does not accelerate graphs that fit in a single GPU's memory but enables processing of graphs that exceed single-GPU capacity.

\subsection{Dynamic Incremental Updates}

We evaluate dynamic edge insertions on 2 and 4 GPUs by comparing the performance of the dynamic update against static recomputation on the modified graph. Table~\ref{tab:dynamic_time} details the speedups for both insertions and deletions across all graphs for fixed and relative batch sizes.

\begin{table*}[htbp]
\centering
\footnotesize
\resizebox{\textwidth}{!}{%
\begin{tabular}{ll rrrrr rrrrr}
\toprule
& & \multicolumn{5}{c}{\textbf{Insertion Speedup}} & \multicolumn{5}{c}{\textbf{Deletion Speedup}} \\ \cmidrule(lr){3-7} \cmidrule(l){8-12}
\textbf{Graph} & \textbf{GPUs} & \textbf{1K} & \textbf{10K} & \textbf{0.1\%} & \textbf{1\%} & \textbf{10\%} & \textbf{1K} & \textbf{10K} & \textbf{0.1\%} & \textbf{1\%} & \textbf{10\%} \\
\midrule
BN1 & 2 & 7.71$\times$ & 5.75$\times$ & 3.82$\times$ & 2.00$\times$ & 0.44$\times$ & 19.70$\times$ & 7.25$\times$ & 1.23$\times$ & 0.61$\times$ & 0.97$\times$ \\
 & 4 & 6.69$\times$ & 5.05$\times$ & 2.80$\times$ & 1.85$\times$ & 0.29$\times$ & 6.83$\times$ & 5.22$\times$ & 1.13$\times$ & 1.03$\times$ & 0.51$\times$ \\ \addlinespace
BN2 & 2 & 3.23$\times$ & 3.75$\times$ & 3.26$\times$ & 2.02$\times$ & 0.49$\times$ & 6.18$\times$ & 2.02$\times$ & 1.78$\times$ & 0.62$\times$ & 0.64$\times$ \\
 & 4 & 2.96$\times$ & 2.39$\times$ & 2.16$\times$ & 1.75$\times$ & 0.36$\times$ & 4.70$\times$ & 4.38$\times$ & 1.64$\times$ & 1.31$\times$ & 1.14$\times$ \\ \addlinespace
CO & 2 & 10.46$\times$ & 10.86$\times$ & 6.96$\times$ & 3.61$\times$ & 0.61$\times$ & 13.75$\times$ & 2.23$\times$ & 1.31$\times$ & 0.93$\times$ & 0.57$\times$ \\
 & 4 & 7.29$\times$ & 5.56$\times$ & 5.36$\times$ & 3.71$\times$ & 0.49$\times$ & 8.75$\times$ & 2.11$\times$ & 2.25$\times$ & 1.87$\times$ & 0.40$\times$ \\ \addlinespace
ER & 2 & 13.08$\times$ & 11.14$\times$ & 6.93$\times$ & 6.81$\times$ & 0.62$\times$ & 13.98$\times$ & 1.95$\times$ & 1.91$\times$ & 0.64$\times$ & 0.28$\times$ \\
 & 4 & 7.23$\times$ & 6.69$\times$ & 6.56$\times$ & 4.33$\times$ & 0.36$\times$ & 9.20$\times$ & 6.65$\times$ & 1.62$\times$ & 1.70$\times$ & 0.32$\times$ \\ \addlinespace
USA & 2 & 10.58$\times$ & 8.10$\times$ & 8.47$\times$ & 7.24$\times$ & 3.07$\times$ & 4.60$\times$ & 6.71$\times$ & 2.97$\times$ & 2.31$\times$ & 1.28$\times$ \\
 & 4 & 9.11$\times$ & 8.66$\times$ & 9.23$\times$ & 8.15$\times$ & 3.61$\times$ & 13.58$\times$ & 5.60$\times$ & 4.93$\times$ & 4.84$\times$ & 3.20$\times$ \\ \addlinespace
CF & 2 & 71.61$\times$ & 73.29$\times$ & 30.92$\times$ & 7.33$\times$ & 1.25$\times$ & 138.98$\times$ & 137.27$\times$ & 49.51$\times$ & 8.62$\times$ & 1.10$\times$ \\
 & 4 & 67.03$\times$ & 63.09$\times$ & 25.88$\times$ & 4.24$\times$ & 0.87$\times$ & 117.37$\times$ & 119.17$\times$ & 33.60$\times$ & 7.61$\times$ & 0.71$\times$ \\ \addlinespace
RMAT1 & 2 & 51.47$\times$ & 49.58$\times$ & 30.93$\times$ & 12.65$\times$ & 2.16$\times$ & 69.16$\times$ & 19.69$\times$ & 15.83$\times$ & 5.74$\times$ & 2.90$\times$ \\
 & 4 & 39.46$\times$ & 38.56$\times$ & 25.76$\times$ & 9.56$\times$ & 1.46$\times$ & 48.91$\times$ & 20.42$\times$ & 6.58$\times$ & 6.17$\times$ & 1.94$\times$ \\ \addlinespace
RMAT2 & 2 & 67.81$\times$ & 67.57$\times$ & 33.76$\times$ & 10.22$\times$ & 1.55$\times$ & 43.53$\times$ & 26.21$\times$ & 17.26$\times$ & 3.88$\times$ & 2.21$\times$ \\
 & 4 & 53.86$\times$ & 52.21$\times$ & 29.90$\times$ & 6.26$\times$ & 1.14$\times$ & 58.02$\times$ & 15.15$\times$ & 8.39$\times$ & 5.51$\times$ & 1.20$\times$ \\ \addlinespace
RMAT3 & 2 & 83.03$\times$ & 82.81$\times$ & 33.02$\times$ & 5.88$\times$ & 1.45$\times$ & 109.29$\times$ & 32.74$\times$ & 23.05$\times$ & 2.97$\times$ & 1.48$\times$ \\
 & 4 & 62.55$\times$ & 63.58$\times$ & 27.49$\times$ & 4.61$\times$ & 0.89$\times$ & 70.28$\times$ & 38.15$\times$ & 9.15$\times$ & 3.19$\times$ & 0.98$\times$ \\
\midrule
\textbf{Mean} & 2 & 20.86$\times$ & 19.61$\times$ & 12.07$\times$ & 5.42$\times$ & 1.05$\times$ & 25.72$\times$ & 9.99$\times$ & 5.56$\times$ & 1.88$\times$ & 1.03$\times$ \\
 & 4 & 16.38$\times$ & 14.66$\times$ & 10.08$\times$ & 4.29$\times$ & 0.76$\times$ & 21.08$\times$ & 11.09$\times$ & 4.36$\times$ & 2.95$\times$ & 0.90$\times$ \\
\bottomrule
\end{tabular}}
\caption{Speedup for dynamic incremental and decremental updates compared to static recomputation.}
\label{tab:dynamic_time}
\end{table*}
\textbf{Fixed batch sizes.}
At fixed batch sizes (1K and 10K edges), the dynamic update cost
remains small across all graphs, while the scratch recomputation cost
varies widely depending on graph size and structure. The speedup is
therefore governed by the recomputation baseline. The RMAT and CF
graphs, which have the highest vertex counts (65--67M) and
correspondingly high recomputation costs, achieve speedups of
51--83$\times$ on 2~GPUs at 1K batch. The remaining graphs, with lower
recomputation baselines, achieve 3--13$\times$.

\textbf{Relative batch sizes.}
At relative batch sizes (0.1\%, 1\%, and 10\% of $|E|$), the absolute
number of modified edges varies by orders of magnitude across graphs.
For USA ($d_{avg}=1$), a relative batch translates to a small absolute
edge count, and each update affects at most one or two neighbors,
bounding cascade depth. Consequently, USA retains a speedup above
$3\times$ even at 10\%.

For graphs with higher average degree (BN1: $d_{avg}=112$, ER:
$d_{avg}=75$), 10\% of $|E|$ translates to tens of millions of edges.
When an inserted edge creates an IN--IN conflict, the revoked vertex's
neighbors must each re-evaluate whether they retain any IN neighbor. In
high-degree graphs, these cascading re-evaluations compound, causing the
dynamic update cost to approach or exceed recomputation cost. On
2~GPUs, the crossover (speedup below 1$\times$) occurs at 10\% for BN1
(0.44$\times$), BN2 (0.49$\times$), CO (0.61$\times$), and ER
(0.62$\times$). The large-scale RMAT and CF graphs remain above
1$\times$ at 10\% on 2~GPUs because their recomputation baseline is
substantially higher.

Speedups on 4~GPUs follow the same trends as on 2~GPUs but are
uniformly lower, reflecting the additional inter-GPU synchronization
overhead per Luby iteration. The crossover at 10\% extends to additional
graphs: RMAT3 drops to 0.89$\times$ and CF to 0.87$\times$ on 4~GPUs,
inversions that do not occur on 2~GPUs, where the same graphs achieve
1.45$\times$ and 1.25$\times$, respectively.

\subsection{Dynamic Decremental Updates}

We evaluate dynamic edge deletions using the same batch sizes and GPU
configurations as for insertions. Table~\ref{tab:dynamic_time} also details the speedups for these deletions.

\textbf{Fixed batch sizes.}
At 1K edges, dynamic deletions achieve higher speedups than insertions
on 7 out of 9 graphs. For instance, on 2~GPUs, CF achieves
$\sim$139$\times$ for deletions versus $\sim$72$\times$ for insertions.
This advantage arises from two factors. First, the deletion dynamic
path is cheaper per edge: deleting an edge requires only a local level
decrement on one endpoint, with no neighbor traversal. Insertions, by
contrast, must detect and resolve IN--IN conflicts, which involves
revoking a vertex and traversing its adjacency list. Second, at small
fixed batch sizes, the scratch recomputation baseline after deletion is
often comparable to or higher than after insertion, further favoring
deletion speedup.

\textbf{Relative batch sizes.}
At larger relative batch sizes, this advantage reverses: insertions
outperform deletions on 8 out of 9 graphs at 0.1\%. Two effects drive
this inversion. First, inserting edges produces a denser modified graph,
raising the scratch recomputation baseline; deleting edges yields a
sparser graph, lowering it. At 1\% and 10\%, the insertion scratch
baseline exceeds the deletion scratch baseline on all 9 graphs,
directly inflating insertion speedups relative to deletion. Second, the
deletion dynamic cost grows with batch size as more OUT vertices
accumulate level decrements that reduce their level to zero, producing
a larger set of UNDECIDED vertices for the Luby re-run.
At 1\% on 2~GPUs, deletions fall below 1$\times$ on BN1
(0.61$\times$), BN2 (0.62$\times$), CO (0.93$\times$), and ER
(0.64$\times$), while insertions at the same batch size remain above
1$\times$ on all four graphs.

On graphs with low average degree, each deletion affects fewer
neighbors, limiting cascade depth. USA ($d_{avg}=1$) retains a speedup
above 1$\times$ across all batch sizes on both 2 and 4~GPUs.

On 4~GPUs, the additional synchronization overhead narrows the advantage
at small batches and widens the deficit at large batches. At 10\%,
dynamic deletion falls below 1$\times$ on BN1 (0.51$\times$), CO
(0.40$\times$), ER (0.32$\times$), CF (0.71$\times$), and RMAT3
(0.98$\times$). Graphs with low average degree (USA, BN2) remain above
1$\times$ at all batch sizes.

\subsection{MIS Quality}

We compare the MIS cardinality after each dynamic update with the MIS computed by the static algorithm on the corresponding modified graph. Across all graphs, batch sizes, and update types, the difference in MIS size is less than 1\% compared to a full recomputation. This variation is expected due to the randomness inherent in Luby's algorithm, as different execution orderings can produce different but equally valid MIS solutions. The dynamic update procedure does not introduce any systematic bias toward producing either smaller or larger MISs.

\section{Conclusion and Future Work}\label{sec:conclusion}

We present DMG-MIS, a dynamic multi-GPU algorithm for maintaining a
Maximal Independent Set under batched edge insertions and deletions. Our work
makes three contributions. First, we introduce a dynamic MIS maintenance
algorithm based on Luby's method that confines re-evaluation to the
neighborhood of affected vertices. Second,
DMG-MIS is the first parallel dynamic MIS algorithm for a multi-GPU setting. 
Third, we optimized the static multi-GPU MIS baseline,
achieving geometric mean speedups of $19.10\times$ on
2~GPUs and $16.91\times$ on 4~GPUs.

For dynamic graphs, batched updates achieve geometric mean speedups of up
to $20.86\times$ (insertions) and $25.72\times$ (deletions) over scratch
recomputation at small batch sizes on 2~GPUs, and $16.38\times$ (insertions) and $21.08\times$ (deletions) on 4~GPUs. 

\textbf{Future Work.} All experiments use PCIe 4.0 interconnects.
Evaluating higher-bandwidth interconnects such as NVLink or NVSwitch may
reveal a different performance profile. The graph partitioning is fixed at
initialization and is not updated after edge mutations; graph-aware
repartitioning strategies such as edge-cut minimization may reduce cross-GPU
communication. Finally, an adaptive policy that selects between dynamic update
and static recomputation based on batch size and graph density would eliminate
the crossover observed at large batch sizes on graphs with higher average degrees.


\clearpage
\section{Appendix}

								\subsection{Pseudocode for Core Subroutines}
\label{sec:pseudocode}

This section provides detailed pseudocode for
the key subroutines referenced in
Algorithms~\ref{algo1:luby_multi}--\ref{alg:delete}.
We begin with the two local computation
procedures, \textsc{mis\-Local} and
\textsc{set\-To\-Out}, which are shared across the
static, insertion, and deletion phases. We then
present the inter-GPU synchronization
procedures. Although the main text refers to
them uniformly as \textsc{sync\-Level} and
\textsc{sync\-Mark}, their semantics differ
across the static Luby loop, the insertion
phase, and the deletion phase. To make these
differences explicit, we use distinct names:
\textsc{sync\-Level\-Luby},
\textsc{sync\-Mark\-Luby},
\textsc{sync\-Level\-Insert},
\textsc{sync\-Mark\-Insert}, and
\textsc{sync\-Mark\-Delete}. The deletion phase
does not require a separate \textsc{sync\-Level}
step, as explained below.

All subroutines rely on an auxiliary Boolean
array $\text{sync}[\cdot]$, replicated across
all devices, which flags vertices whose state
has changed and must be communicated during the
subsequent synchronization phase. This flag is
set during local computation and consumed by
the synchronization procedures.

\subsubsection{misLocal}

\begin{algorithm}[t]
\caption{misLocal(done\_local)}
\label{alg:computemis}
\SetKwProg{Fn}{}{ \{}{\}}
\Fn(){misLocal(done\_local)}{
    \ForEach{local vertex $v \in V_i$ \textbf{in parallel}}{
        \If{$v$ is UNDECIDED}{
            $\text{isLocalMax} = \text{true}$;

            \ForEach{neighbor $u \in \mathcal{N}(v)$}{
                \If{$\pi(u) > \pi(v)$}{
                    \If{$u$ is UNDECIDED}{
                        $\text{done\_local} = \text{false}$; \label{alg4:defer}

                        $\text{isLocalMax} = \text{false}$;

                        \textbf{break}; \Comment{Defer to higher priority}
                    }
                    \If{$u$ is IN}{ \label{alg4:in_start}
                        $\text{mark}[v] = \text{false}$; \label{alg4:in_mark}

                        $\text{sync}[v] = \text{true}$;

                        $\text{isLocalMax} = \text{false}$;

                        \textbf{break}; \Comment{Exclude: IN neighbor}
                    }
                }
            }

            \If{$\text{isLocalMax} == \text{true}$}{ \label{alg4:join_start}
                $\text{level}[v] = \infty$; \Comment{$v$ joins the MIS} \label{alg4:join_level}

                $\text{mark}[v] = \text{true}$;

                $\text{sync}[v] = \text{true}$;
            } \label{alg4:join_end}
        }
    }
}
\end{algorithm}

Algorithm~\ref{alg:computemis} evaluates
each undecided vertex in the local
partition $V_i$. For each such vertex $v$,
the procedure iterates over its neighbors.
If a neighbor $u$ with $\pi(u) > \pi(v)$ is
still undecided, $v$ cannot yet determine
its fate; the convergence flag is set to
false (Line~\ref{alg4:defer}) and the local maximum flag
is cleared. If instead $u$ is already IN,
$v$ immediately transitions to OUT by
setting $\text{mark}[v] = \text{false}$
(Lines~\ref{alg4:in_start}--\ref{alg4:in_mark}). If $v$ survives the entire
neighbor scan (\emph{i.e.},
$\text{isLocalMax}$ remains true), it means
every neighbor either has a lower priority
or has already been decided as OUT. Thus,
$v$ is the local priority maximum and
transitions to IN by setting
$\text{level}[v] = \infty$ (Lines~\ref{alg4:join_start}--\ref{alg4:join_end}).
In all cases where $v$ changes state, the
$\text{sync}$ flag is set to stage the
transition for processing by
$\textsc{setToOut}$ or inter-GPU
synchronization.

\subsubsection{setToOut}

\begin{algorithm}[t]
\caption{setToOut()}
\label{alg:settoout}
\SetKwProg{Fn}{}{ \{}{\}}
\Fn(){setToOut()}{
    \ForEach{local IN vertex $v$ with $\text{sync}[v] == \text{true}$ \textbf{in parallel}}{
        $\text{sync}[v] = \text{false}$;

        \ForEach{neighbor $u \in \mathcal{N}(v)$}{
            $\text{prev} = \text{mark}[u]$;

            $\text{mark}[u] = \text{false}$; \label{alg5:mark}

            \textbf{atomic}($\text{level}[u] = \text{level}[u] + 1$); \label{alg5:level}

            \If{$\text{prev} == \text{true}$ \textbf{or} $u \notin V_i$}{ \label{alg5:sync_cond}
                $\text{sync}[u] = \text{true}$; \Comment{Fresh OUT transition or remote vertex} \label{alg5:sync_set}
            }
        }
    }
}
\end{algorithm}

Algorithm~\ref{alg:settoout} processes
newly decided IN vertices and
propagates their effect to neighbors.
For each new IN vertex $v$ (identified
by $\text{level}[v] = \infty$,
$\text{mark}[v] = \text{true}$, and
$\text{sync}[v] = \text{true}$), the
procedure clears the sync flag and
traverses the adjacency list. For each
neighbor $u$, it unconditionally sets
$\text{mark}[u] = \text{false}$ and
atomically increments $\text{level}[u]$
(Lines~\ref{alg5:mark}--\ref{alg5:level}). The sync flag for $u$ is
set under two conditions (Lines~\ref{alg5:sync_cond}--\ref{alg5:sync_set}):
(i)~$u$'s mark was previously true,
indicating a fresh UNDECIDED-to-OUT
transition that must be broadcast, or
(ii)~$u$ resides on a different GPU ($u
\notin V_i$), ensuring that accumulated
level increments are staged for the
subsequent $\textsc{syncLevel}$ step,
even if $u$ was already OUT.

\subsubsection{syncLevelLuby}

\begin{algorithm}[t]
\caption{syncLevelLuby()}
\label{alg:sync_level_luby}
\SetKwProg{Fn}{}{ \{}{\}}
\Fn(){syncLevelLuby()}{
    \textbf{--- Sender (on GPU $i$) ---}

    \ForEach{remote vertex $v \notin V_i$ \textbf{in parallel}}{
        \If{$\text{sync}[v]$ \textbf{and} $\text{level}[v] > 0$}{
            $\text{val} = \text{level}[v]$;

            \If{$\text{mark}[v] == \text{true}$}{
                $\text{val} = \text{val} + 1$; \Comment{Encode UNDECIDED status}
            }
            Send $(v, \text{val})$ to Owner($v$);

            $\text{sync}[v] = \text{false}$;
            
            $\text{level}[v] = 0$;
        }
    }

    \textbf{--- Receiver (on GPU $j = \text{Owner}(v)$) ---}

    \ForEach{received $(v, \text{val})$ from each sender \textbf{in parallel}}{
        \eIf{$v$ is UNDECIDED}{
            \textbf{atomic}($\text{level}[v] = \text{level}[v] + \text{val}$);

            $\text{mark}[v] = \text{false}$;

            $\text{sync}[v] = \text{true}$;
        }{
            \If{$v$ is OUT}{
                \textbf{atomic}($\text{level}[v] = \text{level}[v] + \text{val}$);
            }
        }
    }
}
\end{algorithm}

Algorithm~\ref{alg:sync_level_luby}
synchronizes $\text{level}$ increments
across GPUs during the static Luby
loop. Each GPU scans its local copies
of remote vertices and, for each vertex
with a set sync flag and a positive
accumulated level, transmits the total
increment to the owning GPU. If
$\text{mark}[v]$ is still true, the
transmitted value is incremented by one
to account for the state transition.
After transmission, the sender resets
$\text{level}[v]$ to zero and clears
the sync flag. On the receiver side, if
$v$ is UNDECIDED, the received value is
atomically added to $\text{level}[v]$
and $v$ transitions to OUT by setting
$\text{mark}[v] = \text{false}$ and
flagging it for broadcast. If $v$ is
already OUT, the level is simply
updated without triggering a broadcast.
The key invariant is that the
UNDECIDED-to-OUT transition for any
vertex occurs exactly once; subsequent
level updates are silent.

\subsubsection{syncMarkLuby}

\begin{algorithm}[t]
\caption{syncMarkLuby()}
\label{alg:sync_mark_luby}
\SetKwProg{Fn}{}{ \{}{\}}
\Fn(){syncMarkLuby()}{
    \textbf{--- Sender (on GPU $i$) ---}

    \ForEach{local vertex $v \in V_i$ \textbf{in parallel}}{
        \If{$\text{sync}[v]$}{
            Broadcast $v$ to all other GPUs;

            $\text{sync}[v] = \text{false}$;
        }
    }

    \textbf{--- Receiver (on GPU $j \neq i$) ---}

    \ForEach{received vertex $v$ from each sender \textbf{in parallel}}{
        \If{$v$ is UNDECIDED on GPU $j$}{
            $\text{mark}[v] = \text{false}$;
        }
    }
}
\end{algorithm}

Algorithm~\ref{alg:sync_mark_luby}
broadcasts OUT transitions to all
GPUs. Unlike
$\textsc{syncLevelLuby}$, which
sends data only to the vertex's
owning GPU, $\textsc{syncMarkLuby}$
broadcasts to \emph{all} other
GPUs, because any GPU may have a
local vertex adjacent to $v$ and
needs to know that $v$ is no longer
UNDECIDED. The vertices flagged
here are those that transitioned to
OUT during either the local
$\textsc{setToOut}$ phase or remotely
via $\textsc{syncLevelLuby}$. On
the receiver side, $\text{mark}[v]$
is set to $\text{false}$ for each
received vertex that is still
UNDECIDED. Vertices already in the
OUT state are silently skipped.

\subsubsection{syncLevelInsert}

\begin{algorithm}[t]
\caption{syncLevelInsert()}
\label{alg:sync_level_insert}
\SetKwProg{Fn}{}{ \{}{\}}
\Fn(){syncLevelInsert()}{
    \textbf{--- Sender (on GPU $i$) ---}

    \ForEach{remote vertex $v \notin V_i$ \textbf{in parallel}}{
        \If{$\text{sync}[v]$}{
            $\text{val} = \text{level}[v]$;

            \If{$\text{mark}[v] == \text{true}$}{
                $\text{val} = \text{val} + 1$;
            }
            Send $(v, \text{val})$ to Owner($v$);

            $\text{sync}[v] = \text{false}$;

            \eIf{$\text{mark}[v] == \text{true}$}{
                $\text{level}[v] = \infty$; \Comment{Restore remote IN state}
            }{
                $\text{level}[v] = 0$;
            }
        }
    }

    \textbf{--- Receiver (on GPU $j = \text{Owner}(v)$) ---}

    \ForEach{received $(v, \text{val})$ from each sender \textbf{in parallel}}{
        \textbf{atomic}($\text{level}[v] = \text{level}[v] - \text{val}$);
    }
}
\end{algorithm}

Algorithm~\ref{alg:sync_level_insert}
synchronizes level
\emph{decrements} during the
insertion phase. This is
fundamentally different from
$\textsc{syncLevelLuby}$, which
transmits \emph{increments}. During
insertions, after the neighbors of
revoked vertices have their levels
decremented, the resulting remote
decrements are accumulated on the
sender's local copy. The value
encoding mirrors the Luby variant.
After transmission, the sender
restores its local copy to the
canonical state: if
$\text{mark}[v]$ is true
(indicating the vertex was
originally IN), the level is
restored to $\infty$; otherwise, it
is reset to zero. This cleanup is
necessary because the insertion
procedure reuses remote copies, and
stale values would corrupt
subsequent phases. On the receiver
side, the received value is
\emph{subtracted} from the vertex's
level, since the insertion protocol
transmits decrements rather than
increments.

\subsubsection{syncMarkInsert}

\begin{algorithm}[t]
\caption{syncMarkInsert()}
\label{alg:sync_mark_insert}
\SetKwProg{Fn}{}{ \{}{\}}
\Fn(){syncMarkInsert()}{
    \textbf{--- Sender (on GPU $i$) ---}

    \ForEach{local vertex $v \in V_i$ \textbf{in parallel}}{
        \If{$\text{sync}[v]$}{
            \eIf{$\text{mark}[v]$}{
                $\text{val} = 1$;
            }{
                $\text{val} = 0$;
            }
            Broadcast $(v, \text{val})$ to all other GPUs;

            $\text{sync}[v] = \text{false}$;
        }
    }

    \textbf{--- Receiver (on GPU $j \neq i$) ---}

    \ForEach{received $(v, \text{val})$ from each sender \textbf{in parallel}}{
        $\text{mark}[v] = \text{val}$;

        $\text{level}[v] = 0$;
    }
}
\end{algorithm}

Algorithm~\ref{alg:sync_mark_insert}
broadcasts the resolved states of
vertices after the insertion phase.
Unlike $\textsc{syncMarkLuby}$,
which only broadcasts OUT
transitions (sending a bare
vertex~ID),
$\textsc{syncMarkInsert}$ transmits
$(v, \text{val})$ pairs where
$\text{val}$ encodes the new
$\text{mark}$ value. This is
necessary because the insertion
phase can produce both OUT
transitions (revoked vertices that
retain at least one IN neighbor)
and UNDECIDED transitions (revoked
vertices or orphaned OUT vertices
whose level has dropped to zero).
On the receiver side, the mark is
directly overwritten and the level
is reset to zero. The level reset
is safe because the owning GPU has
already computed the authoritative
level; the remote copy needs only
the mark to correctly evaluate the
vertex during the subsequent
Luby re-run.

\subsubsection{syncMarkDelete}

\begin{algorithm}[t]
\caption{syncMarkDelete()}
\label{alg:sync_mark_delete}
\SetKwProg{Fn}{}{ \{}{\}}
\Fn(){syncMarkDelete()}{
    \textbf{--- Sender (on GPU $i$) ---}

    \ForEach{local vertex $v \in V_i$ \textbf{in parallel}}{
        \If{$\text{mark}[v] == \text{false}$ \textbf{and} $\text{level}[v] == 0$}{
            $\text{mark}[v] = \text{true}$;

            $\text{level}[v] = 0$; \Comment{To UNDECIDED}

            Broadcast $v$ to all other GPUs;
        }
    }

    \textbf{--- Receiver (on GPU $j \neq i$) ---}

    \ForEach{received vertex $v$ from each sender \textbf{in parallel}}{
        $\text{mark}[v] = \text{true}$;

        $\text{level}[v] = 0$;
    }
}
\end{algorithm}

Algorithm~\ref{alg:sync_mark_delete}
is the sole synchronization
procedure required for the deletion
phase. No separate
$\textsc{syncLevel}$ step is needed
because edge deletions are
processed as directed pairs $(src,
dst)$ where the owning GPU of $src$
directly decrements
$\text{level}[src]$ locally
(Algorithm~\ref{alg:delete},
Line~\ref{algo3:l3}). Since each GPU only
decrements the level of its own
vertices, no cross-GPU level
synchronization is required. The
sender phase fuses state resolution
with broadcast: for each local OUT
vertex whose level has dropped to
zero, the vertex transitions to
UNDECIDED and is broadcast to all
other GPUs. On the receiver side,
$\text{mark}[v]$ is set to
$\text{true}$ and
$\text{level}[v]$ to zero,
restoring the UNDECIDED state on
all remote copies. This is the
inverse of $\textsc{syncMarkLuby}$,
which broadcasts OUT ($\text{mark}
= \text{false}$) transitions;
here, the broadcast conveys
UNDECIDED ($\text{mark} =
\text{true}$) transitions.


\clearpage
\onecolumn
\begingroup
\fontencoding{T1}\fontfamily{lmr}\selectfont
\renewcommand{\bfdefault}{bx}
\fontsize{10}{12}\selectfont
\mathversion{rebuttalLM}
\setcounter{table}{0}
\setcounter{equation}{0}
\newcommand{\AppendixSection}[1]{\par\vskip 1.5em\relax\noindent{\large\bfseries #1\par}\vskip 0.5em\relax}
\newcommand{\AppendixSubsection}[1]{\par\vskip 0.8em\relax\noindent{\normalsize\bfseries #1\par}\vskip 0.35em\relax}
\newcommand{\AppendixTableCaption}[2]{\par\vskip\smallskipamount\noindent\textbf{Table #1.} #2\par\vskip\medskipamount}

\begin{multicols}{2}
\AppendixSection{1 Dataset Characteristics}

\noindent We evaluate the nine graphs summarized in Table 1. The listed abbreviations are used throughout all runtime tables.

{\centering\small
\resizebox{\linewidth}{!}{%
\begin{tabular}{l l r r r r}
\hline
Abbr. & Dataset & $|V|$ & $|E|$ & $d_{\mathrm{avg}}$ & $d_{\mathrm{max}}$ \\
\hline
BN1   & brain networks     & 1.8M  & 202M   & 112 & 758 \\
BN2   & brain networks     & 1.8M  & 67M    & 37  & 1,539 \\
CO    & com-orkut          & 3.1M  & 117M   & 38  & 33,313 \\
ER    & Erd\H{o}s--R\'enyi  & 2.0M  & 150M   & 75  & 215 \\
USA   & roadnetwork-usa    & 23.9M & 28.8M  & 1   & 9 \\
CF    & com-friendster     & 65.6M & 1.8B   & 27  & 5,214 \\
RMAT1 & RMAT               & 67.1M & 268.4M & 4   & 16,343 \\
RMAT2 & RMAT               & 67.1M & 536.8M & 8   & 32,477 \\
RMAT3 & RMAT               & 67.1M & 1.07B  & 16  & 65,102 \\
\hline
\end{tabular}%
}\par}
\AppendixTableCaption{1}{Dataset characteristics.}

\noindent$d_{\mathrm{avg}}$ and $d_{\mathrm{max}}$ denote the reported average and maximum degrees.

\columnbreak

\AppendixSection{2 Absolute GPU Runtimes}

\AppendixSubsection{2.1 Static execution}

{\centering\footnotesize
\resizebox{\linewidth}{!}{%
\begin{tabular}{l r r r r r r}
\hline
Graph & \multicolumn{3}{c}{Static MG-MIS (Ours)} &
        \multicolumn{3}{c}{MG-MIS (Previous)} \\
\cline{2-4}\cline{5-7}
      & 1 GPU & 2 GPUs & 4 GPUs & 1 GPU & 2 GPUs & 4 GPUs \\
\hline
ER    & 2.4  & 3.7  & 7.58  & 3.3  & 88.5   & 230.01 \\
BN1   & 3.4  & 4.3  & 7.54  & 8.8  & 1115   & 1590.2 \\
BN2   & 1.3  & 2.2  & 4.91  & 2.7  & 104.5  & 312.11 \\
CO    & 2.4  & 4    & 7.22  & 3.9  & 135.2  & 224.24 \\
USA   & 4    & 5.8  & 15.95 & 4.7  & 9.3    & 15.33 \\
RMAT1 & 14.9 & 20.8 & 27.5  & 15.2 & 221.5  & 164.73 \\
RMAT2 & 20   & 28.2 & 35.11 & 18.4 & 259.3  & 257.86 \\
RMAT3 & 25.6 & 35.3 & 45.59 & 26.5 & 364.5  & 384.6 \\
CF    & 38.2 & 51.3 & 75.15 & OOM  & OOM    & 1165.58 \\
\hline
\end{tabular}%
}\par}
\AppendixTableCaption{2}{Static GPU execution time: our method and the previous MG-MIS implementation (ms).}
\end{multicols}

\begin{multicols}{2}
\AppendixSubsection{2.2 Dynamic updates}

\noindent Across all dynamic update evaluation tables, dyn denotes dynamic update time and scr denotes full recomputation time on the updated graph. Original denotes the runtime reported for the original graph before the batch update. All execution times are in milliseconds (ms); OOM denotes an out-of-memory outcome. Batch sizes expressed as percentages (0.1\%, 1\%, and 10\%) refer to fractions of the total graph edges ($|E|$), while 1K and 10K denote fixed batches of 1,000 and 10,000 edges inserted into or deleted from the graph.
\end{multicols}

{\centering\small
\resizebox{\textwidth}{!}{%
\begin{tabular}{l r rr rr rr rr rr}
\hline
Graph & Original & \multicolumn{2}{c}{0.1\%} & \multicolumn{2}{c}{1\%} &
\multicolumn{2}{c}{10\%} & \multicolumn{2}{c}{1K} & \multicolumn{2}{c}{10K} \\
\cline{3-12}
 & & dyn & scr & dyn & scr & dyn & scr & dyn & scr & dyn & scr \\
\hline
ER    & 6.66   & 0.96 & 6.65   & 0.90 & 6.13   & 10.36  & 6.47   & 0.48 & 6.28   & 0.57 & 6.35 \\
BN1   & 6.92   & 1.82 & 6.96   & 3.62 & 7.24   & 17.41  & 7.64   & 0.92 & 7.09   & 1.23 & 7.07 \\
BN2   & 3.04   & 0.84 & 2.74   & 1.62 & 3.28   & 7.88   & 3.89   & 0.93 & 3.0    & 0.72 & 2.70 \\
CO    & 7.49   & 1.05 & 7.31   & 2.03 & 7.32   & 12.21  & 7.50   & 0.69 & 7.22   & 0.64 & 6.95 \\
USA   & 10.59  & 1.21 & 10.25  & 1.48 & 10.71  & 4.54   & 13.94  & 0.97 & 10.26  & 1.26 & 10.21 \\
RMAT1 & 75.44  & 2.98 & 74.99  & 5.38 & 76.73  & 26.40  & 93.06  & 1.87 & 74.53  & 1.94 & 74.40 \\
RMAT2 & 104.14 & 3.34 & 103.32 & 8.46 & 107.03 & 61.24  & 132.16 & 2.0  & 102.94 & 2.08 & 103.12 \\
RMAT3 & 137.28 & 4.04 & 136.40 & 13.83 & 141.37 & 112.20 & 174.11 & 2.0  & 135.61 & 2.01 & 135.82 \\
CF    & 217.42 & 7.04 & 217.70 & 30.69 & 224.97 & 208.23 & 261.27 & 3.03 & 216.97 & 2.96 & 216.93 \\
\hline
\end{tabular}%
}\par}
\AppendixTableCaption{3}{DMG-MIS: edge insertion on 2 GPUs (ms).}

{\centering\small
\resizebox{\textwidth}{!}{%
\begin{tabular}{l r rr rr rr rr rr}
\hline
Graph & Original & \multicolumn{2}{c}{0.1\%} & \multicolumn{2}{c}{1\%} &
\multicolumn{2}{c}{10\%} & \multicolumn{2}{c}{1K} & \multicolumn{2}{c}{10K} \\
\cline{3-12}
 & & dyn & scr & dyn & scr & dyn & scr & dyn & scr & dyn & scr \\
\hline
ER    & 10.53  & 1.46 & 9.58   & 2.31 & 10.01  & 23.20  & 8.37   & 1.38 & 9.98   & 1.43 & 9.56 \\
BN1   & 10.13  & 3.42 & 9.58   & 5.55 & 10.28  & 36.55  & 10.75  & 1.40 & 9.37   & 1.84 & 9.29 \\
BN2   & 5.70   & 2.31 & 4.98   & 3.01 & 5.27   & 16.52  & 5.88   & 1.66 & 4.91   & 2.07 & 4.94 \\
CO    & 10.58  & 1.86 & 9.97   & 2.69 & 9.99   & 19.34  & 9.56   & 1.36 & 9.92   & 1.79 & 9.95 \\
USA   & 18.20  & 1.64 & 15.14  & 1.93 & 15.72  & 4.88   & 17.61  & 1.65 & 15.03  & 1.75 & 15.15 \\
RMAT1 & 65.65  & 3.41 & 64.37  & 6.23 & 65.43  & 46.22  & 80.21  & 2.31 & 64.26  & 2.32 & 62.69 \\
RMAT2 & 91.16  & 3.43 & 88.37  & 9.54 & 91.23  & 78.76  & 115.14 & 2.23 & 87.99  & 2.31 & 89.08 \\
RMAT3 & 123.62 & 3.96 & 118.40 & 19.91 & 124.60 & 138.24 & 157.63 & 2.20 & 118.50 & 2.26 & 118.0 \\
CF    & 224.86 & 8.61 & 222.84 & 53.39 & 226.18 & 329.45 & 287.22 & 3.31 & 221.86 & 3.51 & 221.46 \\
\hline
\end{tabular}%
}\par}
\AppendixTableCaption{4}{DMG-MIS: edge insertion on 4 GPUs (ms).}

{\centering\small
\resizebox{\textwidth}{!}{%
\begin{tabular}{l r rr rr rr rr rr}
\hline
Graph & Original & \multicolumn{2}{c}{0.1\%} & \multicolumn{2}{c}{1\%} &
\multicolumn{2}{c}{10\%} & \multicolumn{2}{c}{1K} & \multicolumn{2}{c}{10K} \\
\cline{3-12}
 & & dyn & scr & dyn & scr & dyn & scr & dyn & scr & dyn & scr \\
\hline
ER    & 6.68   & 3.25 & 6.20   & 9.40  & 6.02   & 20.28 & 5.74   & 0.58 & 8.11   & 3.17 & 6.19 \\
BN1   & 7.10   & 5.36 & 6.61   & 10.38 & 6.37   & 6.72  & 6.55   & 0.47 & 9.26   & 1.39 & 10.08 \\
BN2   & 3.14   & 1.48 & 2.64   & 4.36  & 2.69   & 4.88  & 3.10   & 0.44 & 2.72   & 1.35 & 2.73 \\
CO    & 7.27   & 5.34 & 6.97   & 7.48  & 6.94   & 11.36 & 6.47   & 0.51 & 7.01   & 3.20 & 7.12 \\
USA   & 10.96  & 3.58 & 10.64  & 4.60  & 10.61  & 7.98  & 10.24  & 2.32 & 10.68  & 1.60 & 10.73 \\
RMAT1 & 75.49  & 5.28 & 75.63  & 9.81  & 75.21  & 20.46 & 70.65  & 1.44 & 74.61  & 3.94 & 75.59 \\
RMAT2 & 105.29 & 6.59 & 104.29 & 17.97 & 103.16 & 32.91 & 95.40  & 1.52 & 105.12 & 5.30 & 104.35 \\
RMAT3 & 139.27 & 7.98 & 137.74 & 35.09 & 136.20 & 55.95 & 123.56 & 3.17 & 137.98 & 5.29 & 138.65 \\
CF    & 217.40 & 4.37 & 216.36 & 24.53 & 211.35 & 168.48 & 185.71 & 1.56 & 216.81 & 1.58 & 216.88 \\
\hline
\end{tabular}%
}\par}
\AppendixTableCaption{5}{DMG-MIS: edge deletion on 2 GPUs (ms).}

{\centering\small
\resizebox{\textwidth}{!}{%
\begin{tabular}{l r rr rr rr rr rr}
\hline
Graph & Original & \multicolumn{2}{c}{0.1\%} & \multicolumn{2}{c}{1\%} &
\multicolumn{2}{c}{10\%} & \multicolumn{2}{c}{1K} & \multicolumn{2}{c}{10K} \\
\cline{3-12}
 & & dyn & scr & dyn & scr & dyn & scr & dyn & scr & dyn & scr \\
\hline
ER    & 10.44  & 4.98  & 8.08   & 4.88  & 8.28   & 24.77 & 7.88   & 1.07 & 9.84   & 1.45 & 9.64 \\
BN1   & 10.15  & 7.35  & 8.30   & 7.89  & 8.10   & 16.05 & 8.24   & 1.29 & 8.81   & 1.70 & 8.87 \\
BN2   & 5.88   & 2.59  & 4.26   & 3.33  & 4.37   & 3.80  & 4.33   & 0.94 & 4.42   & 0.99 & 4.34 \\
CO    & 10.47  & 4.29  & 9.65   & 4.46  & 8.36   & 22.90 & 9.08   & 0.97 & 8.49   & 4.07 & 8.60 \\
USA   & 17.88  & 3.08  & 15.17  & 3.13  & 15.14  & 4.50  & 14.40  & 1.12 & 15.21  & 2.71 & 15.18 \\
RMAT1 & 65.60  & 8.91  & 62.75  & 10.62 & 62.55  & 23.67 & 59.27  & 1.89 & 62.76  & 5.81 & 63.98 \\
RMAT2 & 91.74  & 13.11 & 86.23  & 14.10 & 86.94  & 41.91 & 81.22  & 1.78 & 87.06  & 4.28 & 87.40 \\
RMAT3 & 123.48 & 14.20 & 119.18 & 21.38 & 117.81 & 90.98 & 109.0  & 2.05 & 118.95 & 7.88 & 119.40 \\
CF    & 225.20 & 6.58  & 221.09 & 28.68 & 218.33 & 272.88 & 192.68 & 1.89 & 221.82 & 1.86 & 221.65 \\
\hline
\end{tabular}%
}\par}
\AppendixTableCaption{6}{DMG-MIS: edge deletion on 4 GPUs (ms).}

\begin{multicols}{2}
\AppendixSection{3 Absolute CPU Runtimes}

\noindent All CPU benchmarks were evaluated on a system with a 12th-generation Intel Core i7-1255U processor (10 physical cores and 12 hardware threads; maximum turbo frequency: 4.70 GHz) and 16 GB of RAM. Runs that exceeded the available system memory are reported as OOM.
\end{multicols}

{\centering\small
\begin{tabular}{l r r}
\hline
Graph & Ours & ECL-MIS \\
\hline
ER    & 17.88  & 46.46 \\
BN1   & 9.22   & 24.16 \\
BN2   & 10.69  & 28.29 \\
CO    & 40.68  & 38.64 \\
USA   & 119.42 & 97.48 \\
RMAT1 & 520.82 & 668.91 \\
RMAT2 & 691.87 & 949.17 \\
RMAT3 & 860.75 & 1415.21 \\
CF    & OOM    & OOM \\
\hline
\end{tabular}\par\vskip\smallskipamount
\textbf{Table 7.} Static CPU execution time (ms).\par\vskip\medskipamount}

{\centering\small
\resizebox{\textwidth}{!}{%
\begin{tabular}{l r rr rr rr rr rr}
\hline
Graph & Original & \multicolumn{2}{c}{0.1\%} & \multicolumn{2}{c}{1\%} &
\multicolumn{2}{c}{10\%} & \multicolumn{2}{c}{1K} & \multicolumn{2}{c}{10K} \\
\cline{3-12}
 & & dyn & scr & dyn & scr & dyn & scr & dyn & scr & dyn & scr \\
\hline
ER    & 17.88  & 1.79  & 107.40 & 5.36   & 108.32 & 36.15   & 112.52 & 0.09 & 108.41 & 0.41 & 108.36 \\
BN1   & 9.22   & 6.73  & 38.27  & 26.13  & 42.83  & 161.97  & 71.66  & 0.28 & 40.13  & 0.71 & 41.87 \\
BN2   & 10.69  & 9.38  & 35.43  & 16.99  & 38.18  & 69.42   & 53.74  & 0.11 & 34.29  & 0.82 & 34.52 \\
CO    & 40.68  & 2.60  & 145.30 & 33.82  & 149.65 & 146.41  & 172.19 & 0.14 & 165.43 & 0.80 & 168.51 \\
USA   & 119.42 & 2.93  & 325.99 & 28.97  & 338.60 & 169.43  & 414.76 & 0.15 & 326.52 & 1.33 & 333.26 \\
RMAT1 & 520.82 & 20.07 & 2651.74 & 127.17 & 2755.84 & 1114.06 & 3270.25 & 0.21 & 2647.44 & 2.60 & 2688.16 \\
RMAT2 & 691.87 & 28.24 & 3870.21 & 271.06 & 4106.58 & 2345.63 & 4855.15 & 0.41 & 3856.35 & 1.58 & 3899.74 \\
RMAT3 & 860.75 & 54.18 & 5886.74 & 486.13 & 5886.24 & 4356.76 & 7043.75 & 0.44 & 5899.38 & 2.67 & 6533.93 \\
CF    & OOM    &       & OOM    &       & OOM    &         & OOM    &      & OOM    &      & OOM \\
\hline
\end{tabular}%
}\par}
\AppendixTableCaption{8}{CPU-DMIS (ours): edge insertion (ms).}

{\centering\small
\resizebox{\textwidth}{!}{%
\begin{tabular}{l r rr rr rr rr rr}
\hline
Graph & Original & \multicolumn{2}{c}{0.1\%} & \multicolumn{2}{c}{1\%} &
\multicolumn{2}{c}{10\%} & \multicolumn{2}{c}{1K} & \multicolumn{2}{c}{10K} \\
\cline{3-12}
 & & dyn & scr & dyn & scr & dyn & scr & dyn & scr & dyn & scr \\
\hline
ER    & 17.88  & 2.21  & 107.63 & 9.84  & 114.22 & 17.20  & 109.57 & 0.04 & 110.65 & 0.33 & 106.97 \\
BN1   & 9.22   & 1.55  & 34.73  & 8.99  & 34.99  & 67.20  & 39.66  & 0.05 & 58.92  & 0.30 & 38.65 \\
BN2   & 10.69  & 2.61  & 33.50  & 10.11 & 32.96  & 18.87  & 34.90  & 0.03 & 32.75  & 3.11 & 33.18 \\
CO    & 40.68  & 1.15  & 138.50 & 6.40  & 138.23 & 29.48  & 153.38 & 0.13 & 151.64 & 0.23 & 205.74 \\
USA   & 119.42 & 6.68  & 277.80 & 22.05 & 272.10 & 89.26  & 269.15 & 0.12 & 294.52 & 0.85 & 271.89 \\
RMAT1 & 520.82 & 6.67  & 2505.99 & 40.12 & 2458.54 & 321.44 & 2322.73 & 0.08 & 2488.32 & 0.46 & 2460.41 \\
RMAT2 & 691.87 & 6.99  & 3774.20 & 64.34 & 3659.02 & 500.41 & 3519.60 & 0.19 & 3679.66 & 0.49 & 3682.33 \\
RMAT3 & 860.75 & 11.29 & 5670.00 & 84.83 & 5512.95 & 855.14 & 5308.24 & 0.17 & 6075.79 & 0.55 & 7990.36 \\
CF    & OOM    &       & OOM    &      & OOM    &        & OOM    &      & OOM    &      & OOM \\
\hline
\end{tabular}%
}\par}
\AppendixTableCaption{9}{CPU-DMIS (ours): edge deletion (ms).}

{\centering\small
\resizebox{\textwidth}{!}{%
\begin{tabular}{l rrrrr rrrrr}
\hline
Graph & \multicolumn{5}{c}{Insertion: full recomputation} &
       \multicolumn{5}{c}{Deletion: full recomputation} \\
\cline{2-6}\cline{7-11}
 & 0.1\% & 1\% & 10\% & 1K & 10K & 0.1\% & 1\% & 10\% & 1K & 10K \\
\hline
ER    & 80.86 & 83.23 & 76.00 & 68.53 & 66.86 & 164.34 & 135.26 & 133.84 & 133.67 & 154.19 \\
BN1   & 43.78 & 44.38 & 35.79 & 36.17 & 50.34 & 126.00 & 125.65 & 127.38 & 131.56 & 125.29 \\
BN2   & 34.56 & 34.64 & 43.16 & 63.21 & 44.17 & 62.32  & 84.85  & 91.76  & 96.03  & 69.82 \\
CO    & 60.25 & 75.90 & 69.92 & 69.71 & 78.08 & 145.54 & 120.31 & 148.50 & 132.13 & 137.21 \\
USA   & 124.05 & 163.01 & 154.66 & 120.37 & 139.44 & 153.73 & 244.01 & 122.61 & 169.11 & 195.21 \\
RMAT1 & 976.20 & 1151.04 & 1365.60 & 1077.44 & 1065.80 & 1158.86 & 1290.60 & 1054.53 & 1117.45 & 1131.53 \\
RMAT2 & 1529.60 & 1455.28 & 2044.73 & 1506.34 & 1542.38 & 1560.16 & 1594.18 & 1523.68 & 1507.65 & 1518.08 \\
RMAT3 & 2206.18 & 2081.74 & 3056.91 & 2146.19 & 1974.22 & 2348.27 & 2069.36 & 2281.13 & 2131.09 & 2127.44 \\
CF    & OOM & OOM & OOM & OOM & OOM & OOM & OOM & OOM & OOM & OOM \\
\hline
\end{tabular}%
}\par}
\AppendixTableCaption{10}{ECL-MIS: full recomputation on updated graphs using the CPU (ms).}

\AppendixSection{4 ECL-MIS Baseline Runtime on an RTX 4090}

{\centering\small
\begin{tabular}{l r}
\hline
Graph & Time (ms) \\
\hline
ER    & 3.34 \\
BN1   & 4.00 \\
BN2   & 2.33 \\
CO    & 1.86 \\
USA   & 2.64 \\
RMAT1 & 19.54 \\
RMAT2 & 26.80 \\
RMAT3 & 35.73 \\
CF    & OOM \\
\hline
\end{tabular}\par\vskip\smallskipamount
\textbf{Table 11.} ECL-MIS baseline execution time on an NVIDIA RTX 4090 (ms).\par\vskip\medskipamount}

\clearpage
\endgroup
\twocolumn


\begin{thebibliography}{00}
													\bibitem{1} G.~E. Blelloch, J.~T. Fineman, and J.~Shun,
														``Greedy sequential maximal independent set and matching are
														parallel on average,''
														in \textit{Proceedings of the 24th ACM Symposium on Parallelism in
														Algorithms and Architectures (SPAA)}, 2012, pp.~308--317.

													\bibitem{2} A.~U. Chaudhry, J.~W. Chinneck, and R.~H.~M. Hafez,
														``Fast heuristics for the frequency channel assignment problem in
														multi-hop wireless networks,''
														\textit{European Journal of Operational Research}, vol.~251, no.~3,
														pp.~771--782, 2016.

												\bibitem{3} J.~M. Byskov,
													``Enumerating maximal
													independent sets with
													applications to graph
													colouring,''
													\textit{Operations Research
													Letters}, vol.~32, no.~6,
													pp.~547--556, 2004.

												\bibitem{4} Y.~Lin, W.~Li, J.~Gu, H.~Ren, B.~Khailany, and D.~Z. Pan,
													``ABCDPlace: Accelerated batch-based concurrent detailed placement on
													multi-threaded CPUs and GPUs,''
													\textit{IEEE Transactions on Computer-Aided Design of Integrated
													Circuits and Systems}, vol.~39, no.~12, pp.~5083--5096, 2020.

												\bibitem{5} M.~Luby, ``A simple
													parallel algorithm for the
													maximal independent set
													problem,'' \textit{SIAM
													Journal on Computing},
													vol.~15, no.~4,
													pp.~1036--1053, 1986.

												\bibitem{6} S.~Assadi, K.~Onak,
													B.~Schieber, and
													S.~Solomon, ``Fully dynamic
													maximal independent set
													with sublinear update
													time,'' in \textit{Proc.\
													50th Annual ACM SIGACT
													Symposium on Theory of
													Computing (STOC)}, 2018,
													pp.~815--826.

												\bibitem{7} S.~Assadi, K.~Onak,
													B.~Schieber, and
													S.~Solomon, ``Fully dynamic
													maximal independent set
													with sublinear in $n$
													update time,'' in
													\textit{Proc.\ 30th Annual
													ACM-SIAM Symposium on
													Discrete Algorithms
													(SODA)}, 2019,
													pp.~1919--1936.

												\bibitem{8} M.~Gupta and
													S.~Khan, ``Simple dynamic
													algorithms for Maximal
													Independent Set and other
													problems,'' in
													\textit{Proc.\ Symposium on
													Simplicity in Algorithms
													(SOSA)}, 2021, pp.~86--91.

												\bibitem{9} S.~Chechik and
													T.~Zhang, ``Fully dynamic
													maximal independent set in
													expected poly-log update
													time,'' in \textit{Proc.\
													60th IEEE Annual Symposium
													on Foundations of Computer
													Science (FOCS)}, 2019,
													pp.~85--100.

												\bibitem{10} A.~Bernstein,
													S.~Forster, and
													M.~Henzinger, ``A
													deamortization approach for
													dynamic spanner and dynamic
													maximal matching,'' in
													\textit{Proc.\ 30th Annual
													ACM-SIAM Symposium on
													Discrete Algorithms
													(SODA)}, 2019,
													pp.~1899--1918.

												\bibitem{11} S.~Behnezhad,
													M.~Derakhshan,
													M.~Hajiaghayi, C.~Stein,
													and M.~Sudan, ``Fully
													dynamic maximal independent
													set with polylogarithmic
													update time,'' in
													\textit{Proc.\ 60th IEEE
													Annual Symposium on
													Foundations of Computer
													Science (FOCS)}, 2019.

												\bibitem{12} G.~E. Blelloch,
													A.~Brady, L.~Dhulipala,
													J.~Fineman, and J.~Lo,
													``Parallel batch-dynamic
													maximal independent set,''
													\textit{arXiv preprint
													arXiv:\allowbreak 2604.07515}, 2026.

												\bibitem{13} K.~Censor-Hillel,
													E.~Haramaty, and Z.~Karnin,
													``Optimal dynamic
													distributed MIS,'' in
													\textit{Proc.\ ACM
													Symposium on Principles of
													Distributed Computing
													(PODC)}, 2016,
													pp.~217--226.

												\bibitem{14} P.~Nijhara,
													A.~Trivedi, and D.~S.
													Banerjee, ``Fast maximal
													independent sets on dynamic
													graphs,'' in \textit{Proc.\
													33rd Euromicro
													International Conference on
													Parallel, Distributed, and
													Network-Based Processing
													(PDP)}, 2025, pp.~78--85.

												\bibitem{15} M.~Burtscher,
													S.~Devale, S.~Azimi,
													J.~Jaiganesh, and
													E.~Powers, ``A high-quality
													and fast maximal
													independent set
													implementation for GPUs,''
													\textit{ACM Trans.\
													Parallel Comput.}, vol.~5,
													no.~2, Article~8, 2018.

												\bibitem{16} P.~Nijhara and
													D.~S. Banerjee, ``TC-MIS:
													Maximal independent set on
													tensor-cores,'' in
													\textit{Proc.\ 18th
													Workshop on General Purpose
													Processing Using GPU
													(GPGPU@ASPLOS)}, 2026.

												\bibitem{17} J.~Nke, S.~Kang, B.~Rees, and C.-H.~Lee,
												``cuMIS: A unified scalable framework for computing maximal independent sets on trillion-edge graphs,''
												in \textit{Proceedings of the 40th ACM International Conference on Supercomputing (ICS '26)},
												2026, pp.~699--713.
												Available: \url{https://doi.org/10.1145/3797905.3800529}

												\bibitem{18} A.~M. Akathoott,
													B.~V. J. Xavier, and
													M.~Burtscher, ``A multi-GPU
													algorithm for computing
													maximal independent sets in
													large graphs,'' in
													\textit{Proc.\ 2025 ACM
													International Conference on
													Supercomputing (ICS)},
													2025.

												\bibitem{19} M.~Winter,
													D.~Mlakar, R.~Zayer, H.-P.
													Seidel, and M.~Steinberger,
													``faimGraph: High
													performance management of
													fully-dynamic graphs under
													tight memory constraints on
													the GPU,'' in
													\textit{Proc.\
													International Conference
													for High Performance
													Computing, Networking,
													Storage and Analysis (SC)},
													2018.

												\bibitem{20} O.~Green and D.~A.
													Bader, ``cuSTINGER:
													Supporting dynamic graph
													algorithms for GPUs,'' in
													\textit{Proc.\ IEEE High
													Performance Extreme
													Computing Conference
													(HPEC)}, 2016.

												\bibitem{21} F.~Busato,
													O.~Green, N.~Bombieri, and
													D.~A. Bader, ``Hornet: An
													efficient data structure
													for dynamic sparse graphs
													and matrices on GPUs,'' in
													\textit{Proc.\ IEEE High
													Performance Extreme
													Computing Conference
													(HPEC)}, 2018.

												\bibitem{22} M.~Sha, Y.~Li,
													B.~He, and K.-L. Tan,
													``Accelerating dynamic
													graph analytics on GPUs,''
													\textit{Proc.\ VLDB
													Endowment}, vol.~11, no.~1,
													pp.~107--120, 2017.

												\bibitem{23} K.~J. Concessao,
													U.~Cheramangalath, M.~J.
													R.~Dev, and R.~Nasre,
													``Meerkat: A framework for
													dynamic graph algorithms on
													GPUs,'' \textit{arXiv
													preprint arXiv:2305.17813},
													2023.

												\bibitem{24} C.~Demetrescu,
													A.~V. Goldberg, and D.~S.
													Johnson, Eds., \textit{The
													Shortest Path Problem:
													Ninth DIMACS Implementation
													Challenge}. DIMACS Book
													Series, AMS, 2009.

												\bibitem{25} R.~A. Rossi and
													N.~K. Ahmed, ``The Network
													Data Repository with
													interactive graph analytics
													and visualization,'' in
													\textit{Proc.\ 29th AAAI
													Conference on Artificial
													Intelligence}, 2015.
													Available:
													\url{https://networkrepository.com}

												\bibitem{26} J.~Leskovec and
													A.~Krevl, ``{SNAP
													Datasets}: {Stanford} Large
													Network Dataset
													Collection,'' Jun.\ 2014.
													Available:
													\url{http://snap.stanford.edu/data}

												\bibitem{27} P.~Erd\H{o}s and
													A.~R\'{e}nyi, ``On the
													evolution of random
													graphs,''
													\textit{Publication of the
													Mathematical Institute of
													the Hungarian Academy of
													Sciences}, vol.~5,
													pp.~17--61, 1960.

											\bibitem{28} D.~Chakrabarti,
										Y.~Zhan, and C.~Faloutsos, ``R-MAT: A
									recursive model for graph mining,'' in
								\textit{Proc.\ SIAM International Conference on
							Data Mining (SDM)}, 2004, pp.~442--446.

\bibitem{29} Y.~Du and H.~Zhang, ``Improved algorithms for
	fully dynamic maximal independent set,'' \textit{arXiv preprint
	arXiv:1804.08908}, 2018.

\bibitem{30} Y.~Wang, Y.~Pan, A.~Davidson, Y.~Wu, C.~Yang,
	L.~Wang, M.~Osama, C.~Yuan, W.~Liu, A.~T. Riffel, and
	J.~D. Owens, ``Gunrock: GPU graph analytics,'' \textit{ACM
	Trans.\ Parallel Comput.}, vol.~4, no.~1, Article~3, 2017.

\bibitem{31} {RAPIDS Development Team}, ``RAPIDS: Libraries for
	end to end GPU data science,'' 2023. Available:
	\url{https://rapids.ai}

\bibitem{32} R.~Yadav, W.~Lee, and A.~Aiken, ``Legate Sparse:
	Distributed sparse computing in Python,'' in
	\textit{Proc.\ International Conference for High Performance
	Computing, Networking, Storage and Analysis (SC)}, 2022.

\bibitem{33} {NVIDIA}, ``NVIDIA Collective Communications
	Library (NCCL),'' 2024. Available:
	\url{https://github.com/NVIDIA/nccl}

\bibitem{34} {NVIDIA}, ``NVSHMEM,'' 2024. Available:
	\url{https://developer.nvidia.com/nvshmem}
\bibitem{35} Ashkiani, Saman and Farach-Colton, Martin and Owens, John D.,
  A Dynamic Hash Table for the GPU,IPDPS 2018,pp. 419-429.

\bibitem{36} P.~Nijhara, A.~H. Singh, A.~Trivedi, and D.~S. Banerjee,
  ``{DyGMIS}: Fast and scalable fully dynamic {MIS} maintenance on large evolving graphs,''
  in \textit{Proceedings of the 55th International Conference on Parallel Processing (ICPP '26)},
  2026, pp.~337--347.

\bibitem{37} A.~Trivedi, P.~Nijhara, and D.~S. Banerjee,
  ``Fast {MIS} on incremental graphs,''
  in \textit{2024 IEEE 31st International Conference on High Performance Computing, Data and Analytics Workshop (HiPCW)},
  2024, pp.~139--140.

\bibitem{38} S.~Hong, S.~K. Kim, T.~Oguntebi, and K.~Olukotun,
  ``Accelerating {CUDA} graph algorithms at maximum warp,''
  in \textit{Proceedings of the 16th ACM SIGPLAN Symposium on Principles and Practice of Parallel Programming (PPoPP)},
  2011, pp.~267--276.
  \end{thebibliography}
\end{document}